\documentclass[twocolumn,prd,unsortedaddress,superscriptaddress]{revtex4-1}
\usepackage{amsmath,amssymb,graphicx,hyperref}
\usepackage{slashed,placeins}

\newcommand{\gae}{\lower 2pt \hbox{$\, \buildrel {\scriptstyle >}\over {\scriptstyle
\sim}\,$}}
\newcommand{\lae}{\lower 2pt \hbox{$\, \buildrel {\scriptstyle <}\over {\scriptstyle
\sim}\,$}}
\newcommand{\e}[1]{\ensuremath{\times 10^{#1}}}

\begin{document}

\title{Non-Standard Models, Solar Neutrinos, and Large $\theta_{13}$}
\author{R. Bonventre}
\affiliation{Department of Physics and Astronomy, University of Pennsylvania, Philadelphia, PA 19104}
\author{A. LaTorre}
\altaffiliation[Present address:]{Department of Physics, University of Chicago, Chicago, IL 60637}
\affiliation{Department of Physics, University of California at Berkeley, Berkeley, CA 94720}

\author{J.R. Klein}
\affiliation{Department of Physics and Astronomy, University of Pennsylvania, Philadelphia, PA 19104}

\author{G.D. Orebi Gann}
\affiliation{Department of Physics, University of California at Berkeley, Berkeley, CA 94720}
\affiliation{Nuclear Science Division, Lawrence Berkeley National Laboratory, Berkeley, CA 94720}

\author{S. Seibert}
\affiliation{Department of Physics and Astronomy, University of Pennsylvania, Philadelphia, PA 19104}

\author{O. Wasalski}
\altaffiliation[Present address:]{Department of Physics and Astronomy, University of British Columbia, Vancouver, BC V6T 1Z1, Canada}
\affiliation{Department of Physics, University of California at Berkeley, Berkeley, CA 94720}

\begin{abstract}

Solar neutrino experiments have yet to see directly the transition region
between matter-enhanced and vacuum oscillations. The transition region is
particularly sensitive to models of non-standard neutrino interactions and
propagation. We examine several such non-standard models, which predict a
lower-energy transition region and a flatter survival probability for the $^8$B
solar neutrinos than the standard large-mixing angle (LMA) model. We find that
while some of the non-standard models provide a better fit to the solar
neutrino data set, the large measured value of $\theta_{13}$ and the size of
the experimental uncertainties lead to a low statistical significance for these
fits. We have also examined whether simple changes to the solar density profile
can lead to a flatter $^8$B survival probability than the LMA prediction, but
find that this is not the case for reasonable changes.  We conclude that the
data in this critical region is still too poor to determine whether any of
these models, or LMA, is the best description of the data.

\end{abstract}

\maketitle

\section{Introduction}

With the recent precision measurements of $\theta_{13}$ \cite{reno,dayabay},
the model of neutrino mixing is nearly complete. Of the seven new parameters
added to the standard model to describe neutrino flavor transformation, only
two remain unmeasured: the sign of the mass difference between the first and
third mass eigenstates, and the value of the CP-violating phase $\delta$. For a
large fraction of neutrino transformation phenomenology, however, the current
knowledge of the parameters is expected to be good enough to describe neutrino
measurements very accurately. Much of the trust in the model comes from the
fact that it neatly mirrors quark mixing, which has been subject to intense
scrutiny for over four decades. Yet the model of neutrino mixing is still just
that---a model---and until we test that model with the kind of precision with
which we have explored the rest of particle physics, we do not know whether it
is in fact a complete description of neutrinos.

Construction of a broad precision measurement program with neutrino
oscillations suffers not only because of the difficulty in detecting neutrinos,
but also because the model makes few predictions other than oscillations
themselves. In vacuum, experiments can measure oscillation behavior very
precisely, but any deviation seen between predicted transformation probability
and observation must first be interpreted as a change to the mixing parameters,
rather than new physics. A search for new physics thus relies primarily on
looking for deviations from the $L/E$ behavior that mass-difference-driven
oscillations must have. Such searches can be sensitive to interesting new
physics scenarios such as transformation to sterile neutrinos
\cite{lsnd,miniboone,minos}, or neutrino decay
\cite{miniboonedecay,bargerdecay}.

The situation is dramatically different once neutrino passage through matter is
considered. The weakness of neutrino interactions allows coherent processes -
including those from new interactions or more exotic physics - to affect flavor
transformation in a measurable way. Indeed, even in Wolfenstein's
\cite{wolfenstein} seminal paper, he considers primarily the effects of
flavor-changing neutral currents (FCNC) as a driver of neutrino flavor
transformation in matter. Mikheyev and Smirnov \cite{ms} subsequently
demonstrated that `standard' oscillations in matter of varying density---such
as that of the Sun---can lead to resonant flavor conversion. This implied that
even tiny effects may be observable. MSW flavor transformation is an explicit
prediction of the Standard Model and the model of neutrino oscillations. It
states that given measured mixing parameters, which can be provided
independently from solar neutrino measurements, and density profiles of the Sun
and the Earth, the phenomenology of the MSW effect is exactly specified.  Yet
any interaction with matter that distinguishes neutrino states, even
interactions weaker than the weak interaction itself, can spoil the agreement
with MSW predictions. That precision measurements using solar neutrinos are
possible has been demonstrated very clearly by the observed hints of non-zero
$\theta_{13}$ that came out of comparing solar neutrino measurements with those
of the KamLAND reactor experiment \cite{snoleta}. The precision of this
comparison rivaled that of the measurements by the dedicated Double CHOOZ
\cite{doublechooz} experiment.

\begin{figure}
  \includegraphics[width=0.5\textwidth]{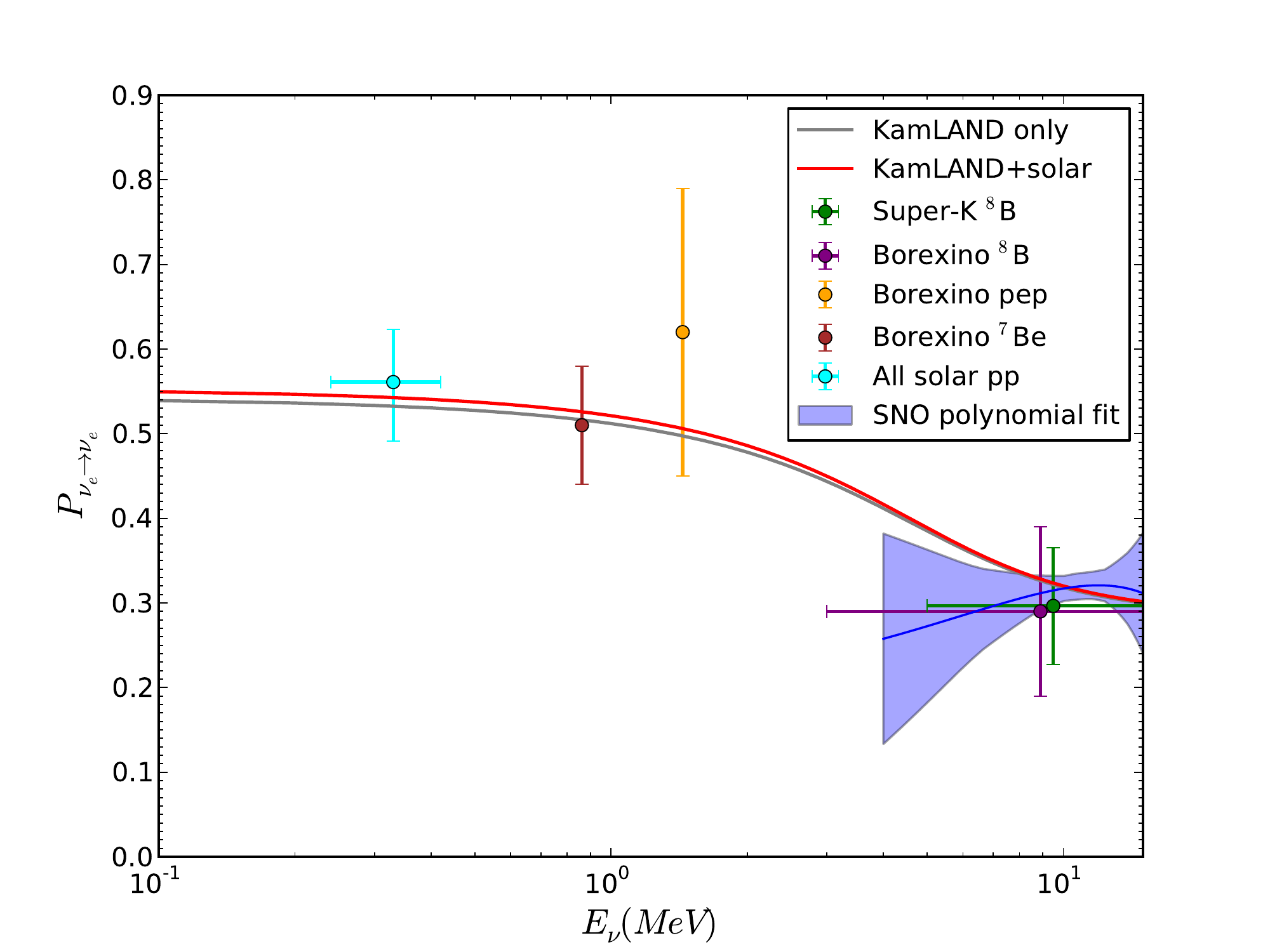}
  \caption{ \label{FIG:mswsurvival} 
    (Color online) MSW prediction for $P_{\nu_{e}\to\nu_{e}}$ for the
    three-flavor KamLAND best fit parameters and the combined solar best fit
    parameters. Note that the $pep$ uncertainties are not Gaussian and the
    value is only $\sim2\sigma$ from zero.  Data points for Borexino and S-K
    $^8$B represent the survival probability averaged over the measured energy
    range.
  }
\end{figure}

While many future experiments \cite{lbne,laguna,hyperk} are planned to
terrestrially observe matter-enhanced oscillations, and thus look for
non-standard effects, to date the only large observed matter enhancement is for
solar neutrinos. In Fig.  \ref{FIG:mswsurvival} we show the predictions of the
survival probability for solar neutrinos, spanning the energy regime from the
lowest-energy $pp$ neutrinos to the highest-energy $hep$ neutrinos. We show
both a curve using just the mixing parameters as measured by KamLAND
\cite{kamland} and one with all solar data included, using the best-fit
large-mixing angle (LMA) parameters. As has been pointed out by many authors
\cite{bahcallregimes,friedlandregimes}, the predicted survival probability has
three regimes. At high energies the effects of matter are pronounced, and thus
the suppression of $\nu_{e}$s exceeds the average value of
$1-1/2\sin^{2}2\theta$ expected for just vacuum oscillations. At low energies
vacuum effects are dominant, thus the survival probability matches the vacuum
value. Between about 1 MeV and 4 MeV there is a transition region between the
low- and high-energy regimes, where the survival probability decreases from the
vacuum average to the matter-dominated value. It is in this transition region
where non-standard effects would be most pronounced, as they interfere with the
expectations from standard MSW transformation. As Nature would have it, probing
this region is particularly difficult. Water Cherenkov experiments have poor
energy resolution and hence difficulty getting below thresholds of 4 MeV,
whereas scintillation experiments are typically either small or restricted to
observing neutrinos through the elastic scattering of electrons, whose
differential cross section is maximally broad.

\begin{figure}
  \includegraphics[width=0.5\textwidth]{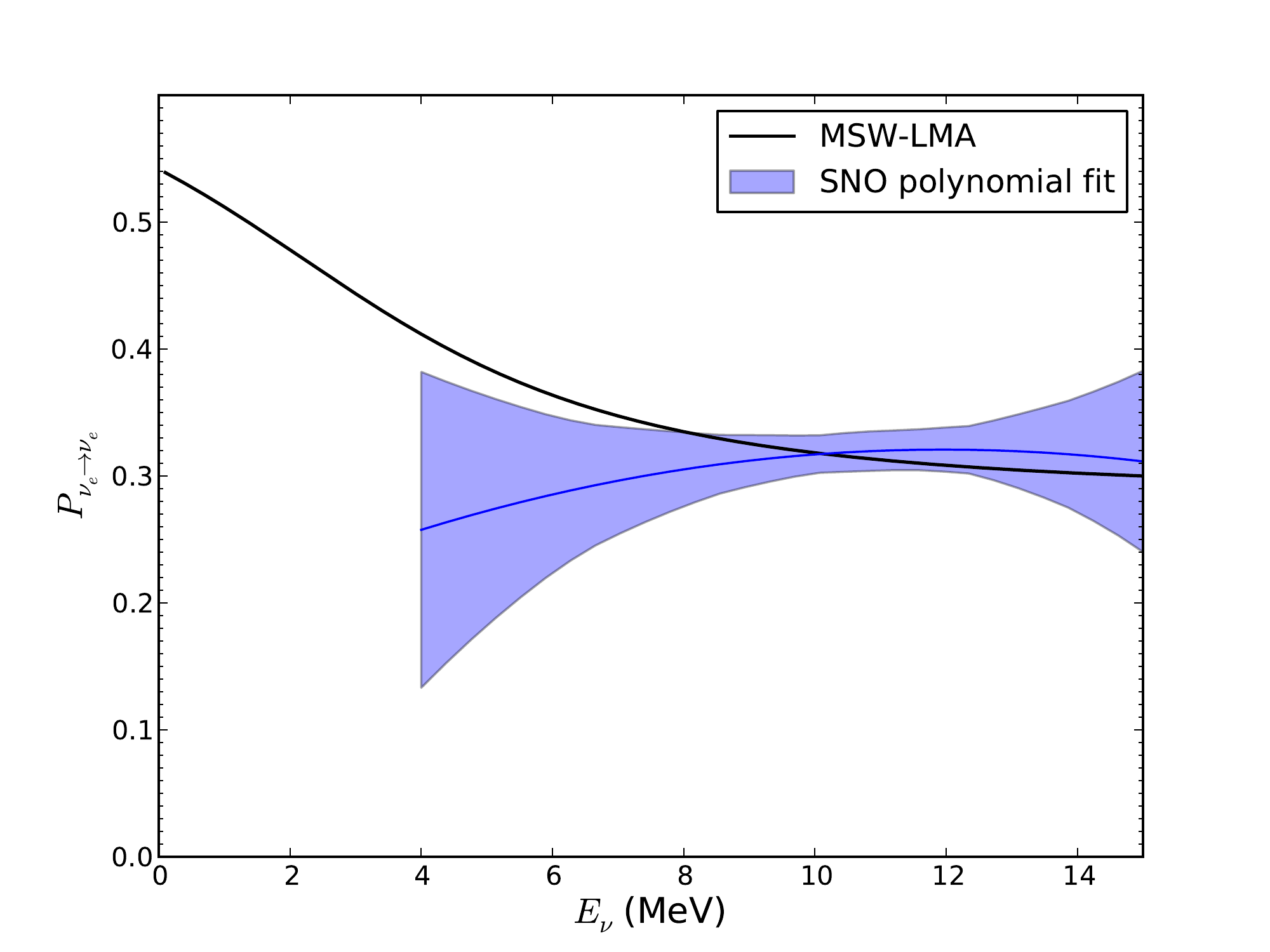}
  \caption{ \label{FIG:mswb8vdatakl}
    (Color online) KamLAND's combined best fit MSW-LMA prediction versus SNO
    extracted $^8B$ survival probability. The band represents the RMS spread at
    any given energy, i.e., not including energy correlations.
  }
\end{figure}

\begin{figure}
  \includegraphics[width=0.5\textwidth]{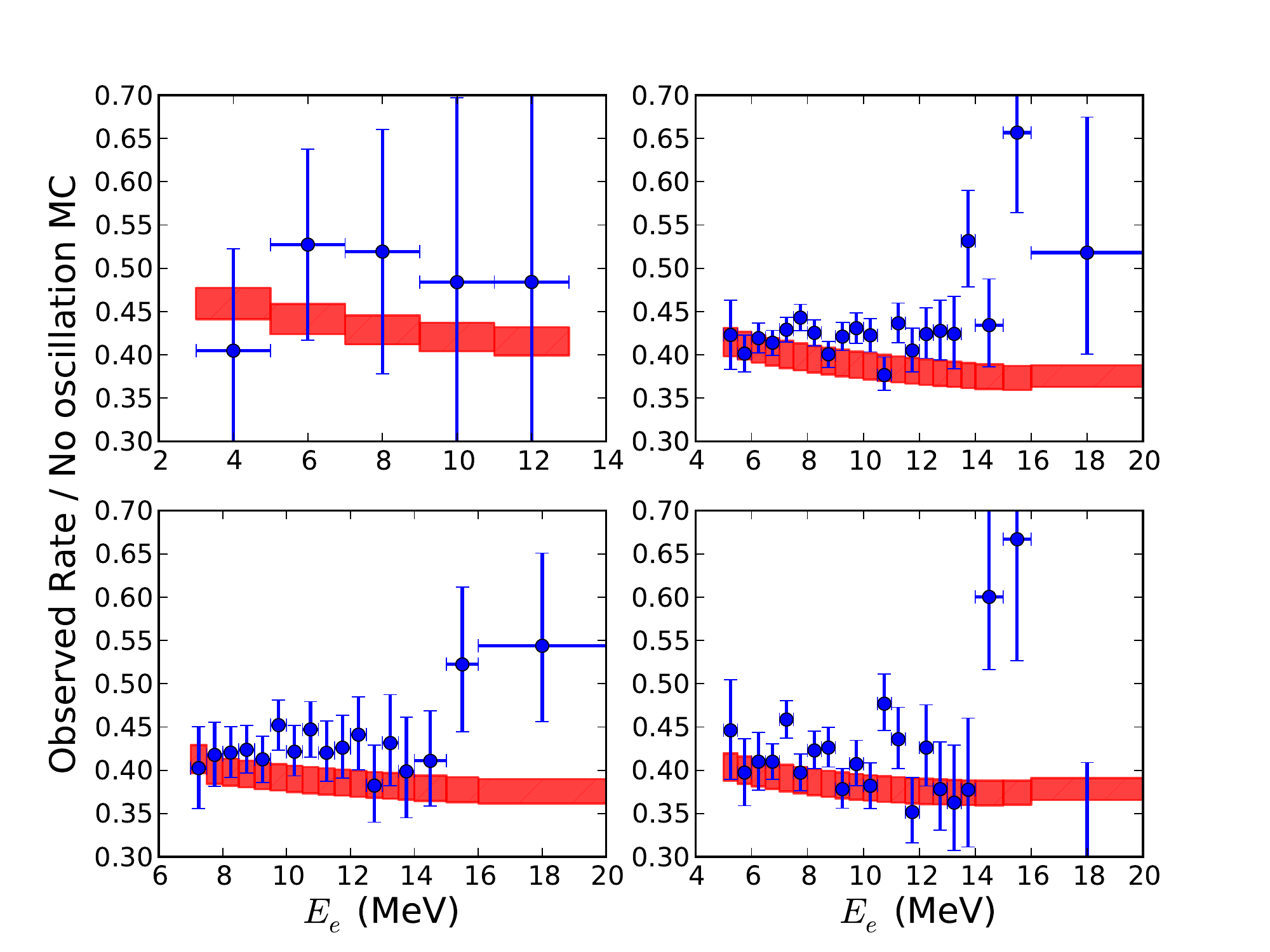}
  \caption{ \label{FIG:superkbxdatakl}
    (Color online) Top left: Borexino, top right: S-K I, bottom left: S-K II,
    bottom right: S-K III. Event rates binned in measured electron energy with
    each bin scaled by Monte Carlo predictions assuming GS98SF2 fluxes, versus
    the same ratio for the expected rates assuming KamLAND's combined best fit
    LMA parameters and SNO's NC $^8B$ flux prediction. Error bars on the data
    points represent statistical and energy uncorrelated systematic
    uncertainties combined in quadrature. Detector response parameters have
    been fixed at their reported value; the width of the band does not include
    the effect of correlated systematic uncertainties. The best fit oscillation
    prediction band width represents the uncertainty on the $^8$B flux. Note
    that we have suppressed the zero for these figures to better illustrate the
    comparison between data and model.
  }
\end{figure}

Many authors
\cite{friedland,holanda,miranda,palazzo,mavanbarger,mavancirelli,mavangonzalez,massivephenomenology,longrange}
have put forth non-standard models and performed fits to the solar neutrino
data set. Prior to the recent $\theta_{13}$ measurements, Palazzo
\cite{palazzo} showed that non-standard interaction models provide a somewhat
better fit to the solar neutrino data than does the standard MSW flavor
transformation. The reason non-standard effects are preferred is the
frustratingly persistent flatness of the high-energy solar $\nu_{e}$ survival
probability, as measured by experiments observing $^8$B neutrinos. In Figs.
\ref{FIG:mswb8vdatakl} and \ref{FIG:superkbxdatakl}, we show the $^8$B
measurements from the Sudbury Neutrino Observatory (SNO), Borexino, and the
Super-Kamiokande (S-K) experiments, with the expectation from large-mixing
angle MSW effect superimposed. We see that while the data is consistent with
MSW, no experiment sees clear evidence of the expected rise due to the matter /
vacuum transition region. The three experiments appear to differ in their
comparison to the model: SNO fits the prediction best at high energies rather
than low, while S-K is the reverse. In other words, SNO's data appears to be
flatter than predicted by MSW due to the fact that at low energies the survival
probability fit is lower than the MSW curve, while S-K's data appears to be
flatter because the high energy event rate is higher than predicted by MSW, but
in all cases the end result is that the data appears flatter than expected.
The Borexino experiment's uncertainties are clearly too large to make a
meaningful comparison with their data alone. 

In this paper we perform fits to the global solar neutrino data sets, including
constraints on $\theta_{13}$ and the most recent measurements by the SNO
collaboration. Section \ref{SEC:datasets} describes each experiment we
consider, how we simulate its results, and how we handle its statistical and
systematic uncertainties. Section \ref{SEC:fit} describes our fitting procedure
and our parameterization of the survival probability for each model we
consider, and the results of the fit for each model are given in Section
\ref{SEC:results}.

\section{Data Sets and Approach}
\label{SEC:datasets}

Our solar neutrino data sets include the weighted average of the results of the
gallium experiments (SAGE, GALLEX, and GNO) given in Ref. \cite{gallium}, and
separately the results of the Chlorine experiment \cite{chlorine}. These
experiments provide integral measurements of several solar neutrino fluxes. For
the `realtime' experiments, which measure exclusive fluxes, we include the most
recent SNO results \cite{sno3phase} for $^8$B, the measurements of S-K I
\cite{sk1}, S-K II \cite{sk2}, and S-K III \cite{sk3} (which are also $^8$B),
and the measurements of Borexino for $^7$Be \cite{borexino7be}, $^8$B
\cite{borexino8b}, and $pep$ \cite{borexinopep}.

We follow the standard approach taken by other authors, except for the handling
of the SNO results, for which explicit energy-dependent survival probabilities
are provided. For all data sets other than SNO we predict the expected number
of events either in a given energy bin or as an integral flux. To achieve this
we convolve the neutrino energy spectrum with its interaction cross section on
a given target, and the outgoing electron energy with the detector's response.
For a given oscillation hypothesis, we include in this integral the energy
dependence of the survival probability. Because of the dependence on the
production region within the Sun we calculate the survival probabilities
separately for each solar neutrino source. The Super-Kamiokande collaboration
has provided bin-by-bin ``no-oscillation'' spectra that include their full
Monte Carlo detector model.  Therefore for a given oscillation hypothesis we
scale their numbers by the ratio of oscillation to no-oscillation calculated
using the analytic Gaussian response they have provided.

Our survival probability calculation is an analytical approximation to a full
three-flavor numerical integration of the wave equation. We assume in all cases
that $\Delta m^{2}_{31}/E$ is much larger than $\Delta m^{2}_{21}/E$ or any
matter potential so the third flavor decouples and propagates independently of
the other two. In addition, we assume adiabatic propagation in the Sun
corrected by a two-flavor jump probability calculated at the resonance of
maximal adiabaticity violation \cite{parke}  (the results agree well with
numerical calculations). We integrate over production location in the Sun for
high metallicity model GS98SF2 \cite{gs98} and low metallicity model AGSS09SF2
\cite{agss09}, using neutrino production and solar density distributions from
each \cite{serenelli}. For the day-night effect we use the procedure described
in Ref. \cite{akhmedov}, modeling the Earth as two spherical shells of constant
density. We use a parameterized average annual solar exposure as described in
Ref. \cite{lisi}. Although we float the mixing parameters in our fits to data,
we constrain them by known terrestrial measurements. For the dominant
$\theta_{12}$ and $\Delta m^{2}_{21}$ parameters we use constraints from
KamLAND \cite{kamland}, and constrain $\theta_{13}$ by the results of the Daya
Bay \cite{dayabay} and RENO \cite{reno} collaborations.

Interaction cross sections for the Chlorine experiment are taken from Bahcall
\cite{b8bahcall}, including the estimated theoretical uncertainties.  For the
Gallium experiments, we assume zero strength for capture to the first two
excited states of $^{71}$Ge, as given in Appendix C of Ref. \cite{gallium} of
the SAGE collaboration. The remaining cross section has uncertainties that are
highly asymmetric for certain energies. We follow Bahcall's suggestion
\cite{bahcallgalliumcs} and take a conservative approach that treats
uncertainties for energies above 2 MeV and uncertainties below 2 MeV as being
correlated with each group but not with each other. To handle the asymmetric
nature of the uncertainties, we use a bifurcated Gaussian. For the elastic
scattering cross section of electrons, which applies to Borexino and
Super-Kamiokande, we use the cross section that includes radiative and
electroweak corrections as given by Bahcall \cite{bahcallcs}.

We consider all experimental uncertainties to be independent, with the
exception of the three S-K measurements for which we treat the normalization
uncertainties as being correlated across the three data sets. We have
marginalized over systematic uncertainties for each experiment. 

For Chlorine, Gallium, and Borexino, we check our reproduction of their data by
comparing their no-oscillation flux predictions to our calculations.  Borexino
only gives a prediction for their integral measurement, but as mentioned
earlier S-K provides binned no-oscillation predictions, allowing us to check
our calculations more carefully. The binned predictions differ from our
calculations by around a few percent per bin, which we assume to be due to
unreported differences between the Gaussian detector response given in Ref.
\cite{sk3} and their full detector Monte Carlo. Once we scale our binned data
by these differences, our integral flux predictions match within one percent.

For the results of the SNO collaboration, we can conveniently use the $\nu_{e}$
survival probability directly. To test a given oscillation hypothesis against
the SNO survival probability, we use the prescription described in Refs.
\cite{snoleta} and \cite{sno3phase}. The survival probability is projected onto
the detected $^8$B spectrum, and the quadratic form used by the SNO
collaboration is extracted. In this way, the comparison comes down to a test of
just six parameters: three for the day survival probability, 
\begin{eqnarray}
  P^{day}_{ee}(E_{\nu} & = & c_0 + c_1(E_{\nu}[\text{MeV}]-10) \nonumber \\
                       &   & + c_2(E_{\nu}[\text{MeV}]-10)^2,
\label{EQ:snopolyday}
\end{eqnarray}
two for the day-night asymmetry,
\begin{equation}
  A_{ee}(E_{\nu} = a_0 + a_1(E_{\nu}[\text{MeV}]-10),
\label{EQ:snopolydaynight}
\end{equation}
and one for the $^8$B flux scale.

\section{Fit}
\label{SEC:fit}

Our interest is in reasonably generic non-standard models, especially those
with the ability to flatten the $^8$B survival probability. For this analysis
we have chosen three types of models: non standard contributions to forward
scattering as described in \cite{friedland}, mass varying neutrinos
\cite{mavanbarger}, and long-range leptonic forces \cite{longrange}.

We used these models to calculate survival probabilities, including the
dominant standard MSW-LMA oscillation. We perform a maximum likelihood fit to
the data, floating the standard mixing parameters ($\theta_{12},\Delta
m^{2}_{21},\theta_{13}$) and various non-standard parameters for each model as
well as the flux scaling for each neutrino production reaction and a systematic
parameter for the shape of the $^8$B spectrum \cite{b8bahcall}. Where we
reference $\chi^2$ in this paper we mean $-2\log\mathcal{L}$. We constrain the
values of the known mixing parameters to the values measured by the KamLAND
collaboration \cite{kamland} for the (1,2) sector, and the measurements of
KamLAND, Daya Bay, and RENO for $\theta_{13}$. The flux for each neutrino
production reaction is constrained by the standard solar model values and
uncertainties, although for $^8$B the main constraint instead comes from SNO's
NC measurement.

\subsection{Non-Standard Forward Scattering}

As suggested by Friedland in \cite{friedland}, one can generically parameterize
these non-standard contributions with an effective low-energy four-fermion
operator
\begin{equation}
  \mathcal{L} = -2\sqrt{2}G_{F}(\bar{\nu}_{\alpha}\gamma_{\rho}\nu_{\beta})(\epsilon^{f\tilde{f}P}_{\alpha\beta}\bar{f}_{P}\gamma^{\rho}\tilde{f}_{P}) + h.c.,
\end{equation}
where P=L,R, and $\epsilon^{f\tilde{f}P}_{\alpha\beta}$ denotes the strength of
the non-standard interaction between neutrinos of flavors $\alpha$ and $\beta$
and the P handed components of fermions $f$ and $\tilde{f}$. Only vector
components where $f = \tilde{f}$ of the non-standard interaction can affect the
neutrino propagation, so we let $\epsilon^{f}_{\alpha\beta} \equiv
\epsilon^{ffL}_{\alpha\beta}+\epsilon^{ffR}_{\alpha\beta}$. One can define
$\epsilon_{\alpha\beta} = \sum_{f=u,d,e}\epsilon^{f}_{\alpha\beta}n_{f}/n_{e}$.
Then the matter part of the generic three flavor NSI oscillation Hamiltonian
can be written as
\begin{equation}
  \mathcal{H} = \sqrt{2}G_{F}n_{e}\begin{pmatrix}
    1+\epsilon_{ee} & \epsilon^{*}_{e\mu} & \epsilon^{*}_{e\tau} \\
    \epsilon_{e\mu} & \epsilon_{\mu\mu} & \epsilon^{*}_{\mu\tau} \\
    \epsilon_{e\tau} & \epsilon_{\mu\tau} & \epsilon_{\tau\tau} \end{pmatrix}.
    \label{EQ:threeflavornsi}
\end{equation}
As in our standard survival probability calculation, we assume the third flavor
decouples and that the non-standard contribution to the potential is much
smaller than $\Delta m^{2}_{31}/E$. Then the effective two flavor Hamiltonian
is
\begin{eqnarray}
  \textbf{H}^{2\nu} =& \frac{\Delta m^{2}_{21}}{4E}\begin{pmatrix}-\cos2\theta_{12} & \sin2\theta_{12} \\ \cos2\theta_{12} & \sin2\theta_{12} \end{pmatrix} \nonumber \\
    & +\sqrt{2}G_{f}n_{e}\begin{pmatrix}\cos\theta_{13} & \epsilon_{1}^{*} \\ \epsilon_{1} & \epsilon_{2} \end{pmatrix}
\end{eqnarray}
where
\begin{eqnarray}
  \epsilon_{1} & = & c_{13}(\epsilon_{e\mu}c_{23}-\epsilon_{e\tau}s_{23}) \nonumber \\
 & & - s_{13}[\epsilon_{\mu\tau}s^{2}_{23}-\epsilon^{*}_{\mu\tau}c^{2}_{23}+(\epsilon_{\mu\mu}-\epsilon_{\tau\tau})c_{23}s_{23}], \\
  \epsilon_{2} & = & \epsilon_{\mu\mu}c^{2}_{23}-(\epsilon_{\mu\tau}+\epsilon^{*}_{\mu\tau})s_{23}c_{23}+\epsilon_{\tau\tau}s^{2}_{23} \nonumber \\
  & & + c^{2}_{13}\epsilon_{ee}+s_{13}[(e^{-i\delta}\epsilon_{e\mu}+e^{i\delta}\epsilon^{*}_{e\mu})s_{23}c_{13} \nonumber \\
  & & + (e^{-i\delta}\epsilon_{e\tau}+e^{i\delta}\epsilon^{*}_{e\tau})c_{13}c_{23}] \nonumber \\
   & & - s^{2}_{13}[\epsilon_{\mu\mu}s^{2}_{23}+(\epsilon_{\mu\tau}+\epsilon^{*}_{\mu\tau})s_{23}c_{23}+\epsilon_{\tau\tau}c^{2}_{23}].
\end{eqnarray}
We follow the example of Ref. \cite{friedland} to calculate a modified mixing
angle in matter as well as a jump probability to get a predicted survival
probability.

\begin{figure}
  \includegraphics[width=0.5\textwidth]{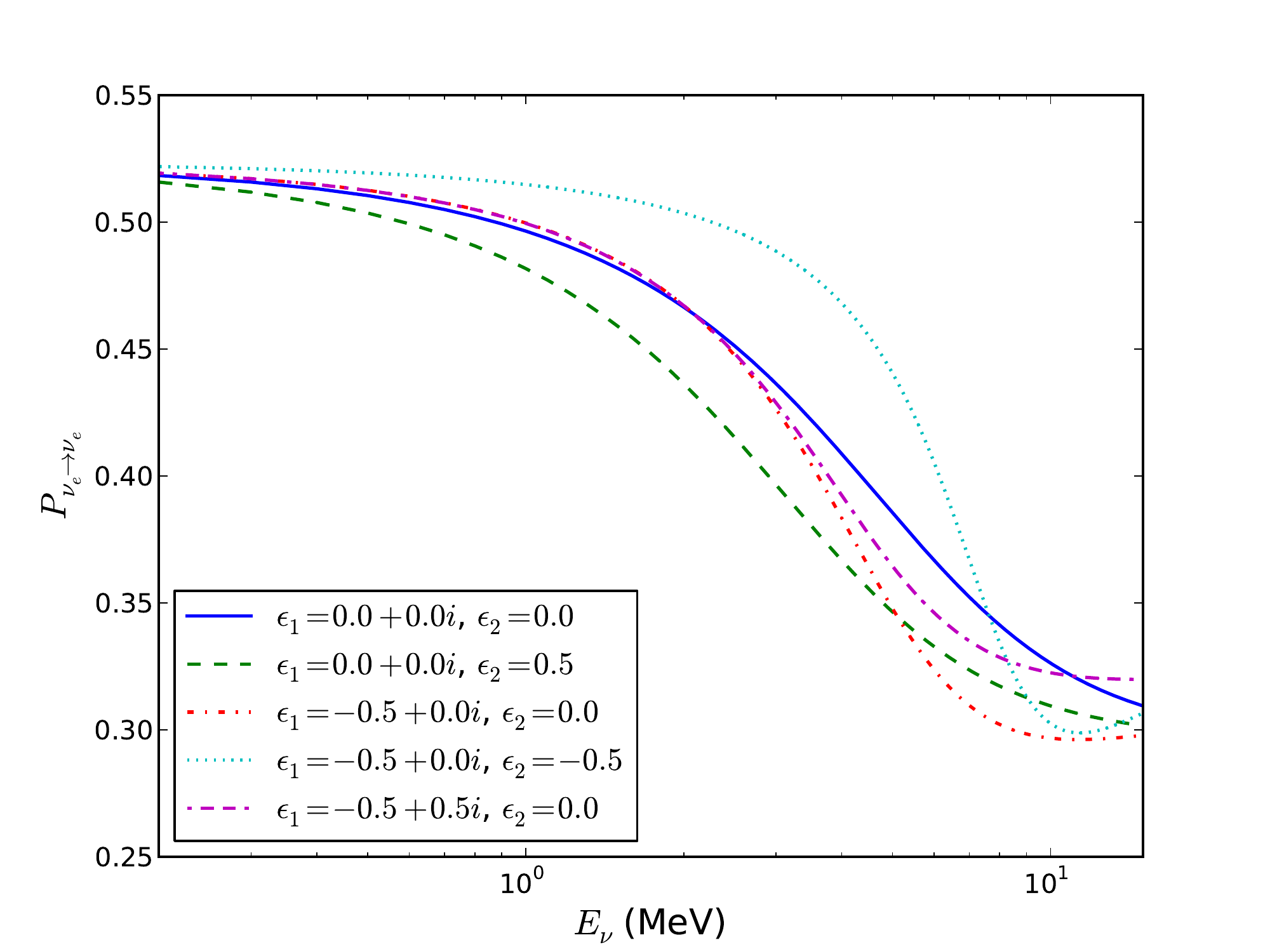}
  \caption{ \label{FIG:nsirange}
    (Color online) Survival probabilities for a range of the NSI parameters
    $\epsilon_{1},\epsilon_{2}$
  }
\end{figure}

This model adds up to three new parameters to the survival probability:
$Re[\epsilon_{1}],Im[\epsilon_{1}],\epsilon_{2}$. Fig. \ref{FIG:nsirange} shows
the effect of each one on the shape of the survival probability.

Current constraints on the strength of these vertices come from accelerator
experiments like NuTeV and CHARM, atmospheric neutrino and charged lepton
experiments like LEP, and by limits on the charged lepton operators. The
parameters $\epsilon_{e\mu},\epsilon_{\mu\mu}$ are well constrained ($\lae
10^{-2}-10^{-3}$), and analysis of atmospheric neutrino data has shown
$\epsilon_{\mu\tau} \lae 10^{-2}$ \cite{massivephenomenology}. However there
remain vertices that can still be quite large, for example,
$|\epsilon^f_{e\tau,ee}| \lae 0.5$, or $|\epsilon^{dR}_{tt}| < 6$.

By letting all the muon vertices go to zero, we get
\begin{eqnarray}
  \epsilon_{1} & = & -c_{13}s_{23}\epsilon_{e\tau} + s_{13}c_{23}s_{23}\epsilon_{\tau\tau}, \\
  \epsilon_{2} & = & s^2_{23}\epsilon_{\tau\tau}+c^2_{13}\epsilon_{ee}+s_{13}c_{13}c_{23}(e^{-i\delta}\epsilon_{e\tau}+e^{i\delta}\epsilon_{e\tau}^{*}) \nonumber \\
               &   &-s^2_{13}c^2_{23}\epsilon_{\tau\tau}.
\end{eqnarray}
The effect of these non-standard parameters on the survival probability as a
function of energy is shown in Fig. \ref{FIG:nsirange}.

\subsection{Mass Varying Neutrinos}

\subsubsection{Neutrino Density Effects}

In Ref. \cite{mavanbarger} it was proposed that neutrinos are coupled to dark
energy in a way that their energy densities track each other. This model was
made to resolve the coincidence of the energy density of dark energy and matter
being similar today even though their ratio scales as $\sim$1/(scale
factor)$^3$. In general this implies so-called `Mass Varying Neutrinos'
(MaVaNs), where the neutrino mass becomes a function of the neutrino density.
If the neutrino couples to a scalar field, then following Ref.
\cite{mavancirelli} at low energy one can write an effective Lagrangian in a
model independent way
\begin{align}
  \textbf{L}(m_i) = & \sum_i \left[ m_i \overline{\nu}^{c}_{i} \nu_{i} + m_i n_i^{C\nu B}\right. \nonumber \\
               & \left. +\int \frac{d^{3}k}{(2\pi)^{3}}\sqrt{k^{2}+m_i^{2}}f_i(k)+V_0(m_i)\right].
\end{align}
Here $n_i^{C\nu B} = 112$ cm$^{-3}$ is the number density of non-relativistic
relic neutrinos of each type and $f_i(k)$ is the occupation number for momentum
k of non-relic neutrinos in our medium (in this case a function of the neutrino
production profile in the Sun). Then one can parameterize the scalar potential
$V_0(m_i) \propto f(m_i/\mu)$ where $\mu$ is some arbitrary mass scale. The
observed equation of state for dark energy implies that the potential must be
flat, while minimizing the total potential implies it must decrease with
increasing neutrino mass. Various forms for the scalar potential have been
suggested, for example, $\log(\mu/m_i)$ or $(m_i/\mu)^{-\alpha}$. For either of
these forms, minimizing the effective potential implies that
\begin{equation}
  m_i(r) \approx m_{i,0} - |U_{e,i}|^{2} A(r) m_{i,0}^2,
\end{equation}
where $m_{i,0}$ is the vacuum mass of $\nu_i$ and
\begin{equation}
  A(r) = \frac{1}{n^{C\nu B}} \int\frac{d^3k}{(2\pi)^3}\frac{1}{\sqrt{k^2+m_i^2}}f_e(k,r).
\end{equation}
Here we have used the fact that $f_i(k,r) = |U_{e,i}|^2f_e(k,r)$
\cite{mavancirelli}.

\begin{figure}
  \includegraphics[width=0.5\textwidth]{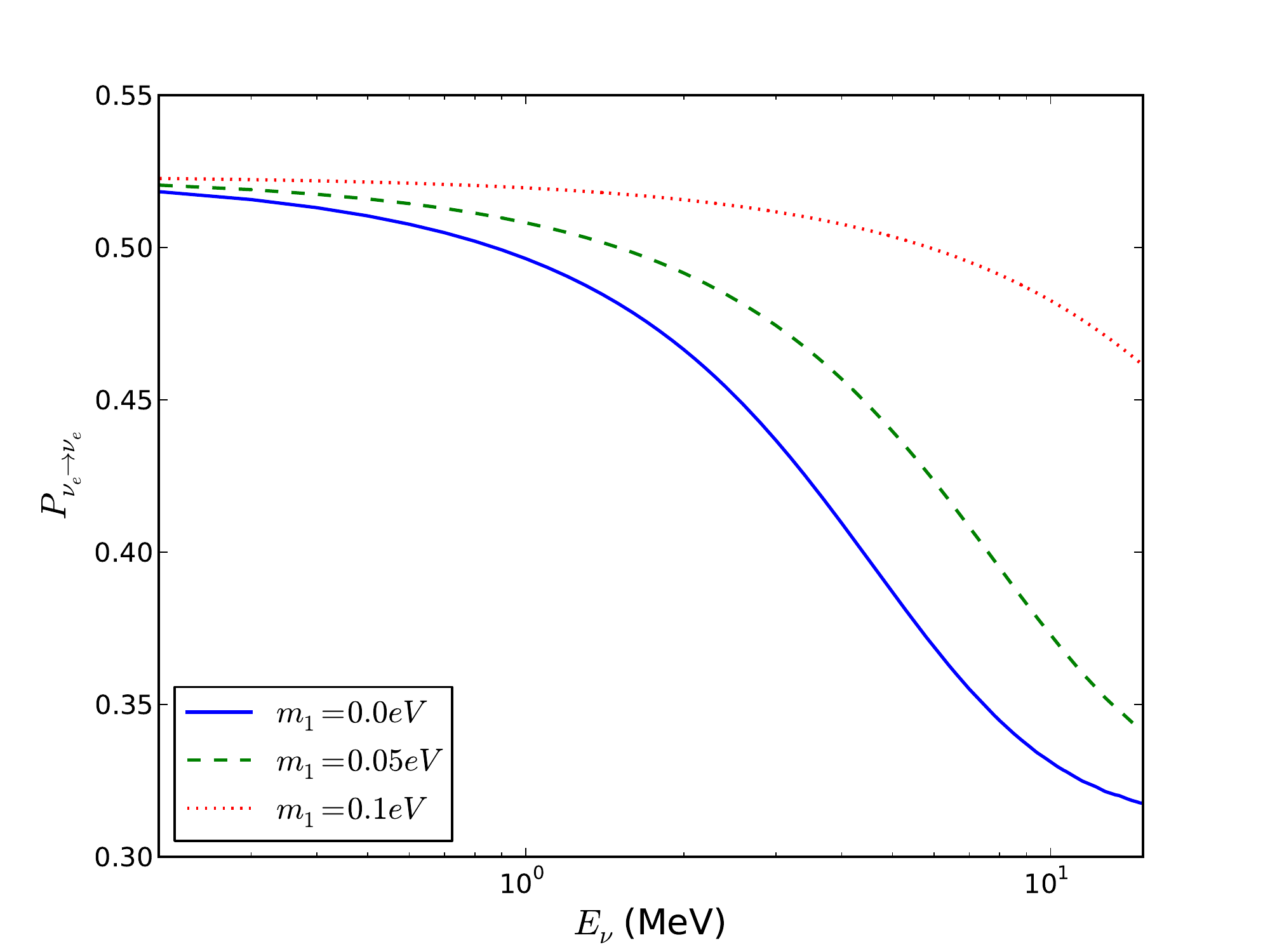}
  \caption{ \label{FIG:mavan1range}
    (Color online) Survival probabilities for the neutrino density dependent
    MaVaN model at several values of $m_{1,0}$
  }
\end{figure}

Then before MSW matter effects, we have
\begin{eqnarray}
  \Delta m^{2}_{21,eff}(r) & = & m_2^2(r)-m_1^2(r) \nonumber \\
                           & \approx & \Delta m^{2}_{21,0} \left[1-3s^2_{12}c^2_{13}A(r) m_{1,0}\right] \nonumber \\
                           &   & +2c^2_{13}A(r)\left[c^2_{12}-s^2_{12}\right]m_{1,0}^{3},
\end{eqnarray}

and we can solve for the survival probability by substituting this effective
mass squared difference into the survival probability calculations for normal
MSW oscillations. Then given a particular distribution of neutrinos, our
effective mass squared difference becomes a function of the vacuum neutrino
mass $m_{1,0}$. The survival probability for various values of the vacuum mass
is shown in Fig. \ref{FIG:mavan1range}.

A previous two-flavor oscillation analysis of solar data and KamLAND found a
$3\sigma$ upper limit of $m_{1,0} < 0.009$ eV, with no improvement in the fit
to the data over MSW-LMA \cite{mavancirelli}.

\subsubsection{Fermion Density Effects}

In addition to the effect described above, it is possible for this scalar field
to couple to visible matter. Ref. \cite{mavangonzalez} parameterizes this model
by adding a light scalar field $\phi$ of mass $m_\phi$, which is weakly coupled
to neutrinos and fermions;

\begin{eqnarray}
  \textbf{L} &=& \sum_i \overline{\nu}_i(i\slashed{\partial}-m_{i,0})\nu_i + \sum_f \overline{f}(i\slashed{\partial}-m_{f,0})f \nonumber \\
             & & + \frac{1}{2}\phi(\partial^2-m_\phi^2)\phi + \sum_{ij}\lambda^{ij}\overline{\nu}_i\nu_j\phi \nonumber \\
             & & + \sum_f \lambda^{f} \overline{f}f\phi.
\end{eqnarray}
Then the elements of the mass matrix become
\begin{eqnarray}
  m_{ij}(r) &=& m_{i,0}\delta_{ij}-M_{ij}(r) \nonumber, \\
  M_{ij}(r) &=& \frac{\lambda^{ij}}{m_\phi^2} \left(\sum_f \lambda^f n_f(r)\right. \nonumber \\
            & & \left.+ \sum_i \lambda^{ii} \int \frac{d^3k}{(2\pi)^3}\frac{M_{ii}}{\sqrt{k^2+M^2_{ii}}}f_i(k,r)\right).
\end{eqnarray}
We will only consider the added effect of the coupling to fermionic matter by
letting $m_{1,0} \sim 0$, such that
\begin{equation}
  M_{ij}(r) = \frac{\lambda^{ij}}{m_\phi^2} \sum_f \lambda^f n_f(r).
  \label{EQ:mavanfermi}
\end{equation}
Assuming that effect of this coupling is small compared to $m_{3,0}$, we can
decouple the third neutrino state. Then diagonalizing the 1-2 sector for the
mass eigenstates in matter gives
\begin{equation}
\cos2\theta^m_{12}(r_0) = \frac{2\Delta m^2_{21}(r)\cos2\theta_{12}-A(r)}{\Delta m^{2}_m}
\label{EQ:mavan2cos2tm}
\end{equation}
where 
\begin{eqnarray}
  \Delta m^{2}_m &=& \left((\Delta m^{2}_{21}(r))^2+4M_3^4(r)\right. \nonumber \\
                 & & \left.-2A(r)\Delta m^2_{21}(r)\cos2\theta_{12}+A^2\right)^{\frac{1}{2}},\\
  \Delta m^{2}_{21}(r) &=& (m_{2,0}-M_2(r))^2-(m_{1,0}-M_1(r))^2,
\end{eqnarray}
and $M_{1,2,3}$ are linear combinations of the $M_{ij}$s, and can be
parameterized as
\begin{equation}
  M_i(r) = \alpha_i \rho(r)
\end{equation}
for matter density $\rho(r)$. Then we can substitute the mixing angle in matter
from Eq. \ref{EQ:mavan2cos2tm} into our standard oscillation equations to get a
survival probability as a function of our parameters $\alpha_i$.

\begin{figure}
  \includegraphics[width=0.5\textwidth]{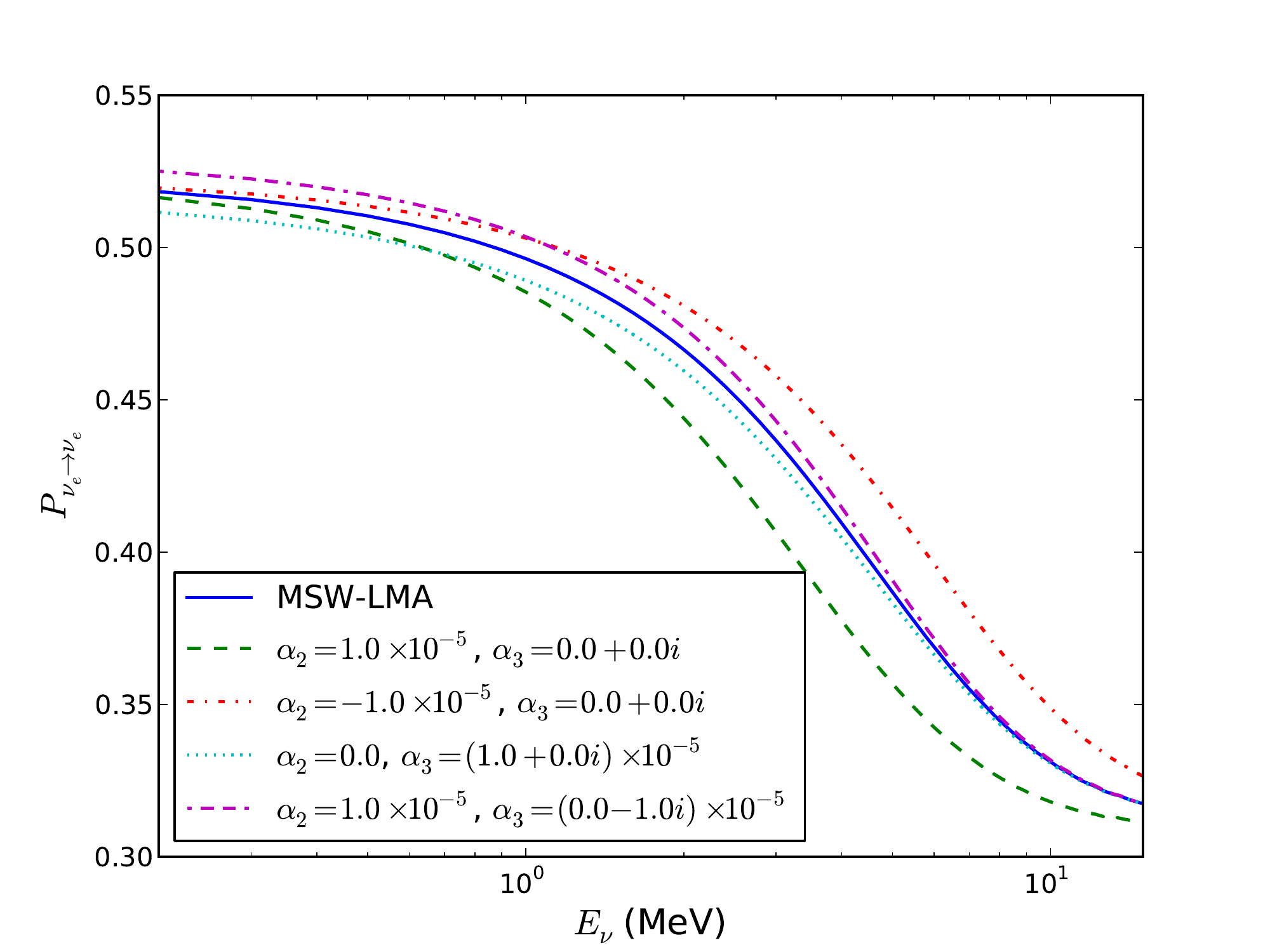}
  \caption{ \label{FIG:mavan2range}
    (Color online) Survival probabilities for the fermion density dependent
    MaVaN model at several values of $\alpha_{2},\alpha_{3}$
  }
\end{figure}

For the KamLAND constraint, we replace $\theta_{12}$ with $\theta_{12}^m$ and
$\Delta m^{2}_{21}$ with $\Delta m^{2}_m$ as defined above except with $A \to
-A$ and $\rho \sim 3$gr/cm$^3$ for the density of the Earth's crust. The
survival probability for various values of the parameters $\alpha_i$ is shown
in Fig. \ref{FIG:mavan2range}.

Current limits for the effective Yukawa coupling of any scalar with $m_\phi
\gae 10^{-11}$eV to nucleons from tests of the inverse square law are
$|\lambda^N| \lae 10^{-21}$ [22]. A previous two-flavor oscillation analysis of
solar data plus KamLAND \cite{mavangonzalez} found 90\% confidence level bounds
of
\begin{eqnarray}
  -2.2\e{-5} \leq &\alpha_2/\text{eV}& \leq 1.4\e{-4}, \\
                  &|\alpha_3|/\text{eV}& \leq 2.3\e{-5} \text{ for } \alpha_3^2 > 0, \\
                  &|\alpha_3|/\text{eV}& \leq 3.4\e{-5} \text{ for } \alpha_3^2 < 0. 
\end{eqnarray}

\subsection{Long-Range Leptonic Forces}

We consider another group of generic non-standard interactions characterized by
a new long-range force coupling to lepton flavor number. Since lepton flavor
number is not conserved, such a force is likely to have a finite range. In
general if the range is long enough, we follow Ref.  \cite{longrange} and write
the effect of the force at some point in the Sun in terms of a function
\begin{equation}
  W(r) = \frac{2\pi\lambda}{r}\int_0^{R_{sun}} dr^{\prime} r^{\prime} n_e(r^{\prime}) \left(e^{-|r^{\prime}-r|/\lambda}-e^{-(r^{\prime}+r)/\lambda}\right),
\end{equation}
where $\lambda$ is the range of the force. Long range forces of this kind can
be probed by studying experimental tests of the equivalence principle; this
sort of analysis was used to get a bound on a vector long-range force's
dimensionless coupling constant $k_V < 10^{-49}$ \cite{okun}. More recently
Gonzalez-Garcia et al \cite{longrange} performed a two flavor oscillation
analysis of solar data to find $3\sigma$ bounds for scalar, vector, and tensor
forces of infinite range that couple to electron number of

\begin{eqnarray}
  k_S(e) &\leq& 5.0\e{-45}\text{ }(m_1=0\text{eV}), \\
  k_S(e) &\leq& 1.5\e{-46}\text{ }(m_1=0.1\text{eV}), \\
  k_V(e) &\leq& 2.5\e{-53}, \\
  k_T(e) &\leq& 1.7\e{-60}\text{eV}^{-1}.
\end{eqnarray}

\subsubsection{Scalar Interaction}

In the case where the new long-range force is a scalar coupling, we see a
similar situation to the MaVaN. We now have a light scalar that only couples to
neutrinos and electrons, which one can parameterize in terms of the function
W(r). The new term in the Lagrangian for the neutrinos is
\begin{equation}
  \textbf{L} = -g_0\phi\overline{\nu}\nu
\end{equation}
and so the kinetic part of the Hamiltonian gains a term
\begin{equation}
  \textbf{M}^{\prime} = U_{12}^{\dagger}U_{13}^{\dagger}U_{23}^{\dagger}
\begin{pmatrix} -M_s(r) & 0 & 0 \\ 0 & 0 & 0 \\ 0 & 0 & 0 \end{pmatrix}
  U_{23}U_{13}U_{12},
\end{equation}
where $M_s(r) = k_s(e)W(r)$ and $k_s(e)=\frac{g_0^2}{4\pi}$. After decoupling
the third flavor and diagonalizing the mass matrix for the remaining two we get
the matter mixing angle in the adiabatic limit of
\begin{equation}
  \sin2\theta^m_{12}(r_0) = \frac{\sin2\theta_{12}\Delta m^2_s}{\Delta m^{\prime 2}_s}
\end{equation}
where
\begin{eqnarray}
  \Delta m^2_s & = & \Delta m^2_{12}-M_s(r_0)\Delta m_{12}c^2_{13}, \\
  (\Delta m^{\prime 2}_s)^2 & = & \left[\Delta m^2_s\cos2\theta_{12}-2E_\nu V(r_0)c^2_{13}\right. \nonumber \\
                            &   & \left.-M_s^2(r_0)c^2_{13}+M_s(r_0)(m_1+m_2)\right]^2 \nonumber \\
                            &   & +\sin2\theta_{12}\Delta m^2_s.
\end{eqnarray}

\begin{figure}
  \includegraphics[width=0.5\textwidth]{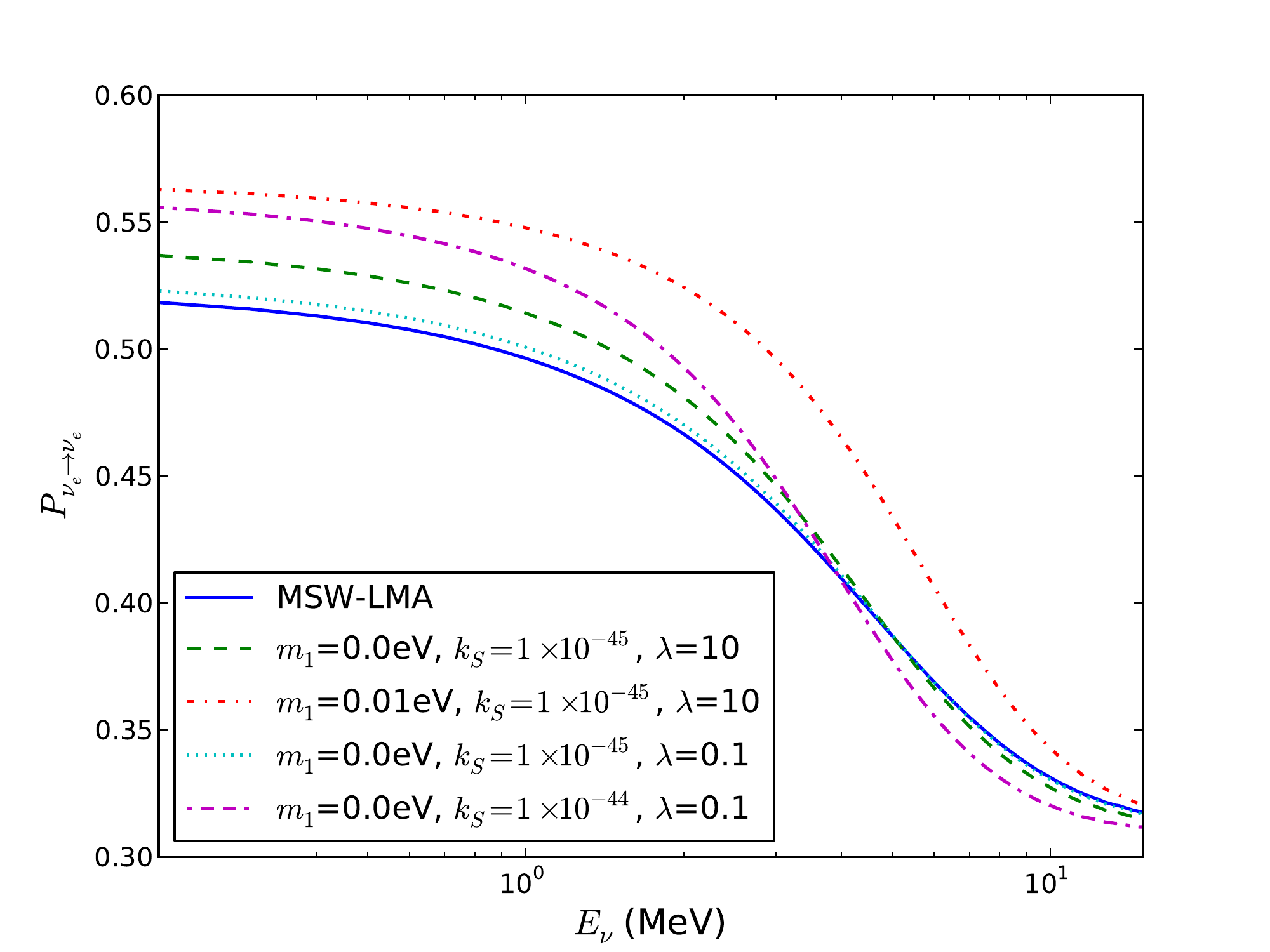}
  \caption{ \label{FIG:lrscalarrange}
    (Color online) Survival probabilities for a long-range scalar interaction
    at various values of the range and strength of the coupling and the
    neutrino mass scale.
  }
\end{figure}

The survival probability for various values of the range and coupling strength
is shown in Fig. \ref{FIG:lrscalarrange}.

\subsubsection{Vector Interaction}

If the force is mediated by a vector boson $A_\alpha$, then 

\begin{equation}
  \textbf{L} = -g_1A_\alpha\overline{\nu}\gamma^{\alpha}\nu
\end{equation}
and the potential $V(r) = V_{MSW} + k_V W(r)$ where $k_V=\frac{g_1^2}{4\pi}$.
We can solve for the survival probability using the standard MSW oscillation
equations, substituting in the above for the MSW potential.

\begin{figure}
  \includegraphics[width=0.5\textwidth]{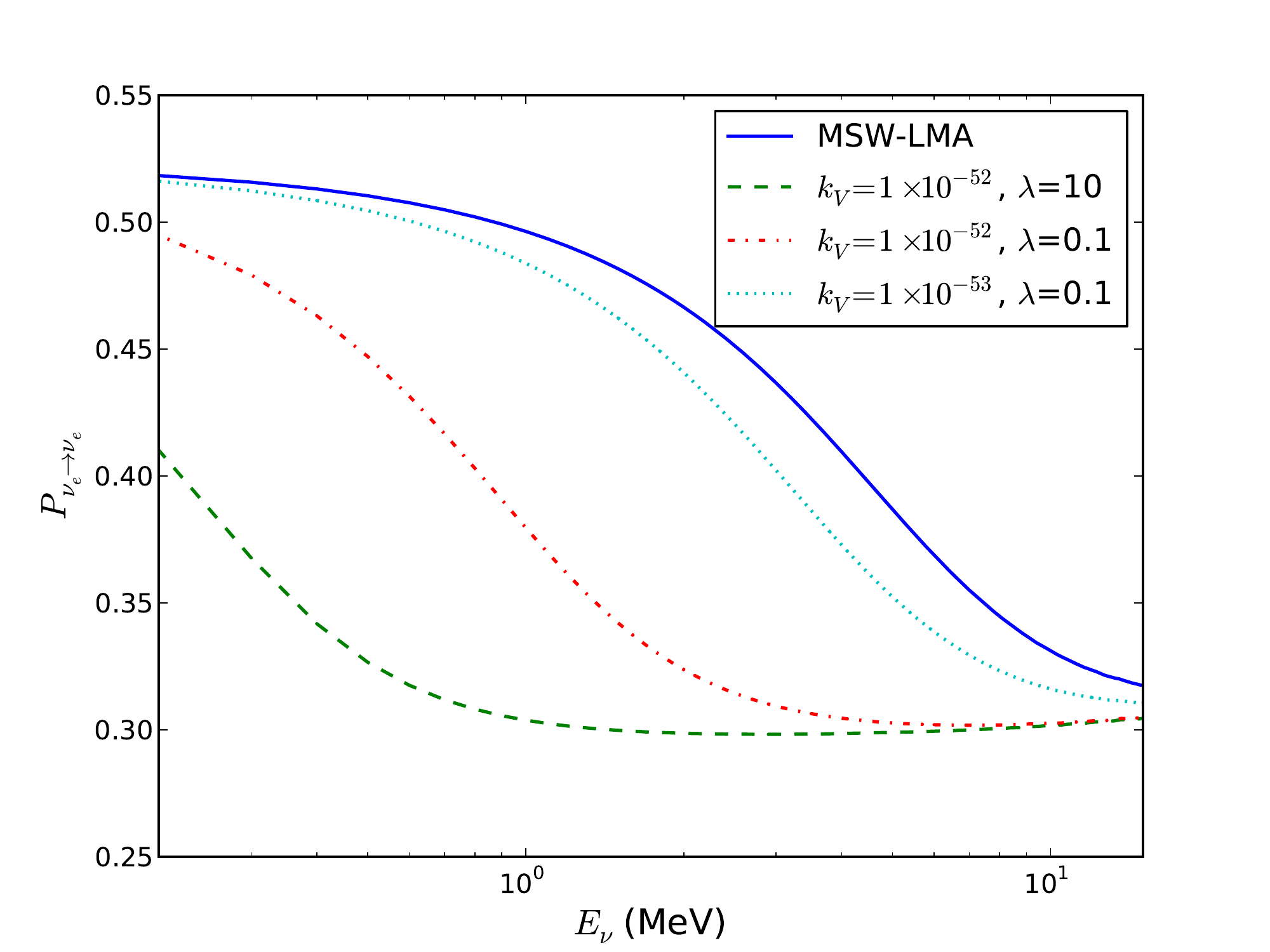}
  \caption{ \label{FIG:lrvectorrange}
    (Color online) Survival probabilities for a long-range vector interaction
    at various values of the range and strength of the coupling.
  }
\end{figure}

The survival probability for various values of the range and coupling strength
is shown in Fig. \ref{FIG:lrvectorrange}.

\subsubsection{Tensor Interaction}

If the force is mediated by a tensor field with spin 2, $\chi_{\alpha\beta}$,
then
\begin{equation}
  \textbf{L} = -\frac{g_2}{2}\chi_{\alpha\beta}\left(\overline{\nu}\gamma^\alpha i\partial^\beta\nu - i\partial^\alpha\overline{\nu}\gamma^\beta\nu\right).
\end{equation}
Now the potential is $V(r) = V_{MSW} + E_\nu k_T W(r)$, where
$k_T=m_e\frac{g_2^2}{4\pi}$. Again we can use the standard MSW oscillation
equations substituting in this new potential.

\begin{figure}
  \includegraphics[width=0.5\textwidth]{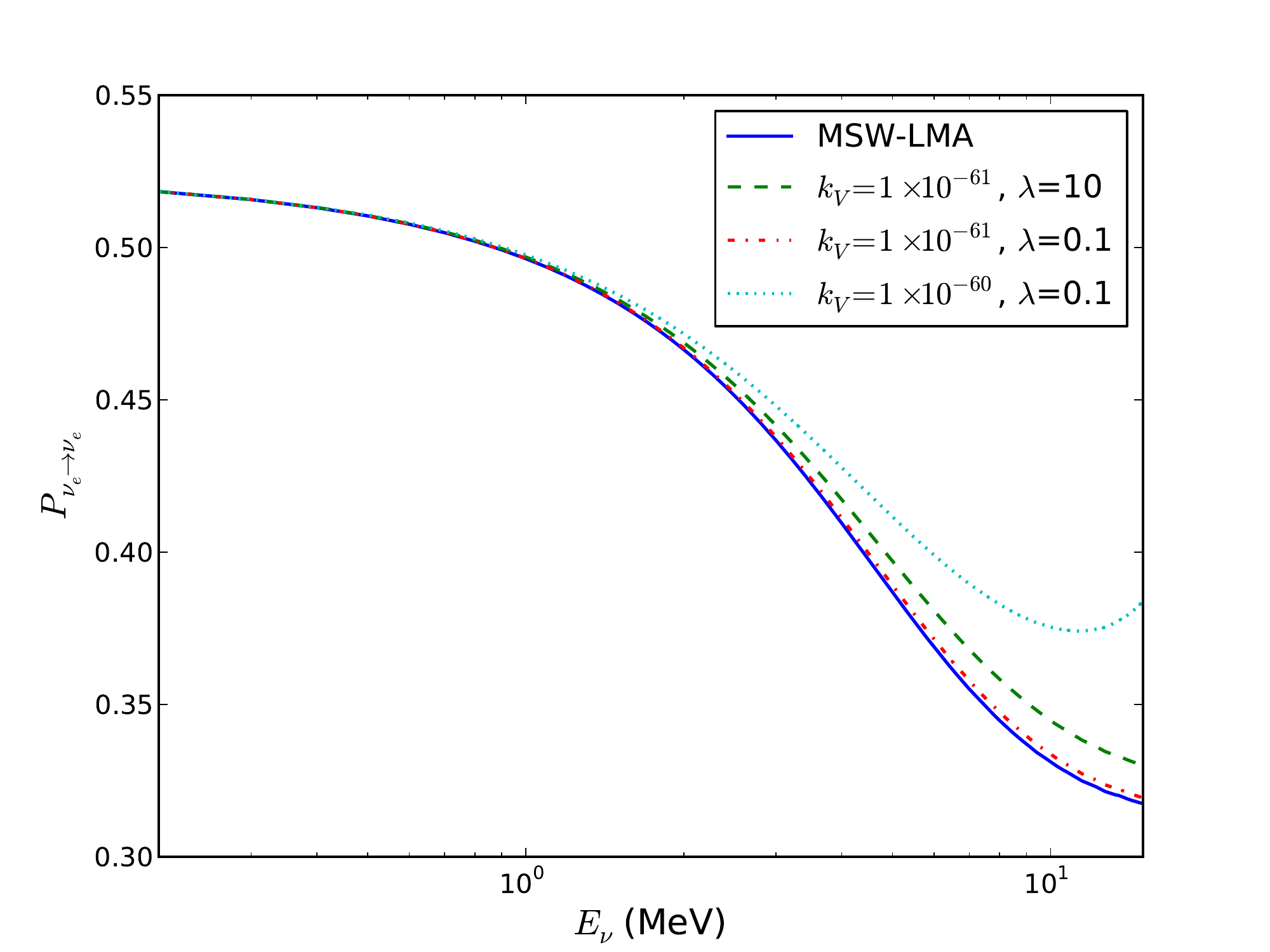}
  \caption{ \label{FIG:lrtensorrange}
    (Color online) Survival probabilities for a long-range tensor interaction
    at various values of the range and strength of the coupling.
  }
\end{figure}

The survival probability for various values of the range and coupling strength
is shown in Fig. \ref{FIG:lrtensorrange}.

\subsection{Non-Standard Solar Model}
\label{SEC:nssolarmodel}

We want to check that any improvement in the fit achieved by replacing MSW with
a non-standard model cannot be easily reproduced by modifying solar model
parameters. In addition, we want to see that we are sensitive to the transition
region independent of exact knowledge of the Sun --- that is, that small
changes in the parameters of the solar model do not create changes in the
transition region on the order of the small effects expected from non-standard
models. To this end, in addition to comparing fits using both the high
metallicity and low metallicity solar models, we use the fact that in the
adiabatic approximation, there are only two inputs from the solar model that
affect the survival probability. They are the absolute flux constraints, and
the convolution of the density profile with the neutrino production profiles.
We can effectively remove many of our assumptions about the solar model from
our fit by removing the absolute flux constraints entirely, and for the other
two sets of parameters, distorting the density profile linearly, so that
\begin{equation}
  n_e^\prime(r) = (1 + \delta_0 + \alpha r)n_e(r)
\end{equation}
for some change in the core density $\delta_0$, where $\alpha$ is determined by
$\delta_0$ and the constraint that the total mass remains the same. A recent
study has shown that a change in the central density is plausible, and was able
to create a model with the central density increased by over 10\% using stellar
evolution software \cite{solarcoredensity}.

We can get a reasonable constraint on the uncertainty of the solar density
profile by comparing the predictions of standard solar models to
helioseismological measurements of the sound profile, which differ by around
1\% \cite{bahcall1000,bahcallhelio}.

In this fit we will not constrain the density change since we are also using it
as a proxy for any change in the production profile. Additionally, although we
cannot use the flux constraints from the solar model in this fit since they are
no longer valid once we change the density, we can constrain the sum of the
fluxes using the luminosity of the Sun \cite{bahcalllumi} and constrain the
ratio of the $pp$ to $pep$ fluxes since the nuclear matrix elements are the
same \cite{pppepratio}.

\section{Results}
\label{SEC:results}

\subsection{Large Mixing Angle MSW}

\begin{figure}
  \includegraphics[width=0.5\textwidth]{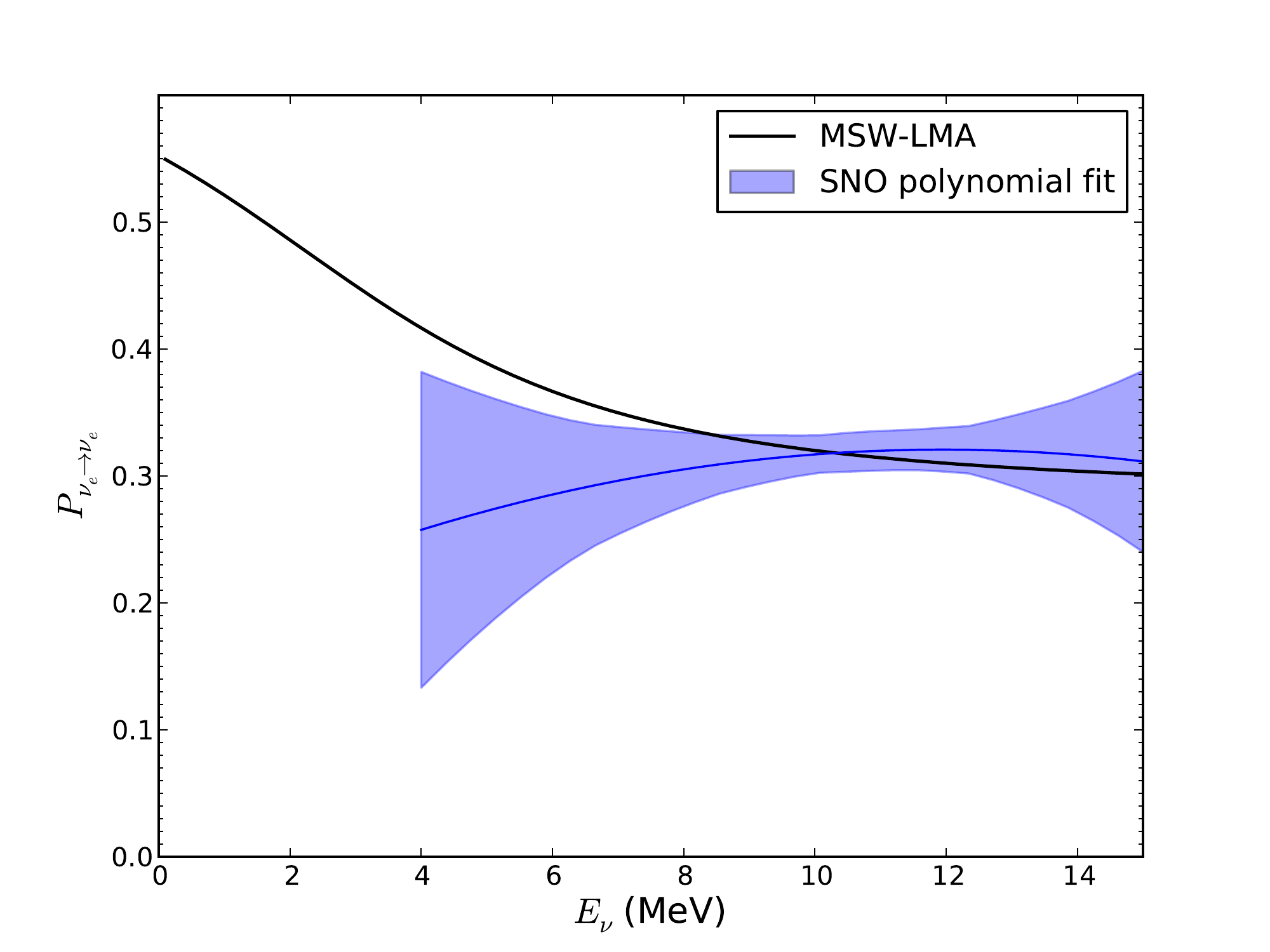}
  \caption{ \label{FIG:mswb8vdata}
    (Color online) Our best fit MSW-LMA prediction versus SNO extracted $^8B$
    survival probability. The band represents the RMS spread at any given
    energy, i.e., not including energy correlations.
  }
\end{figure}

\begin{figure}
  \includegraphics[width=0.5\textwidth]{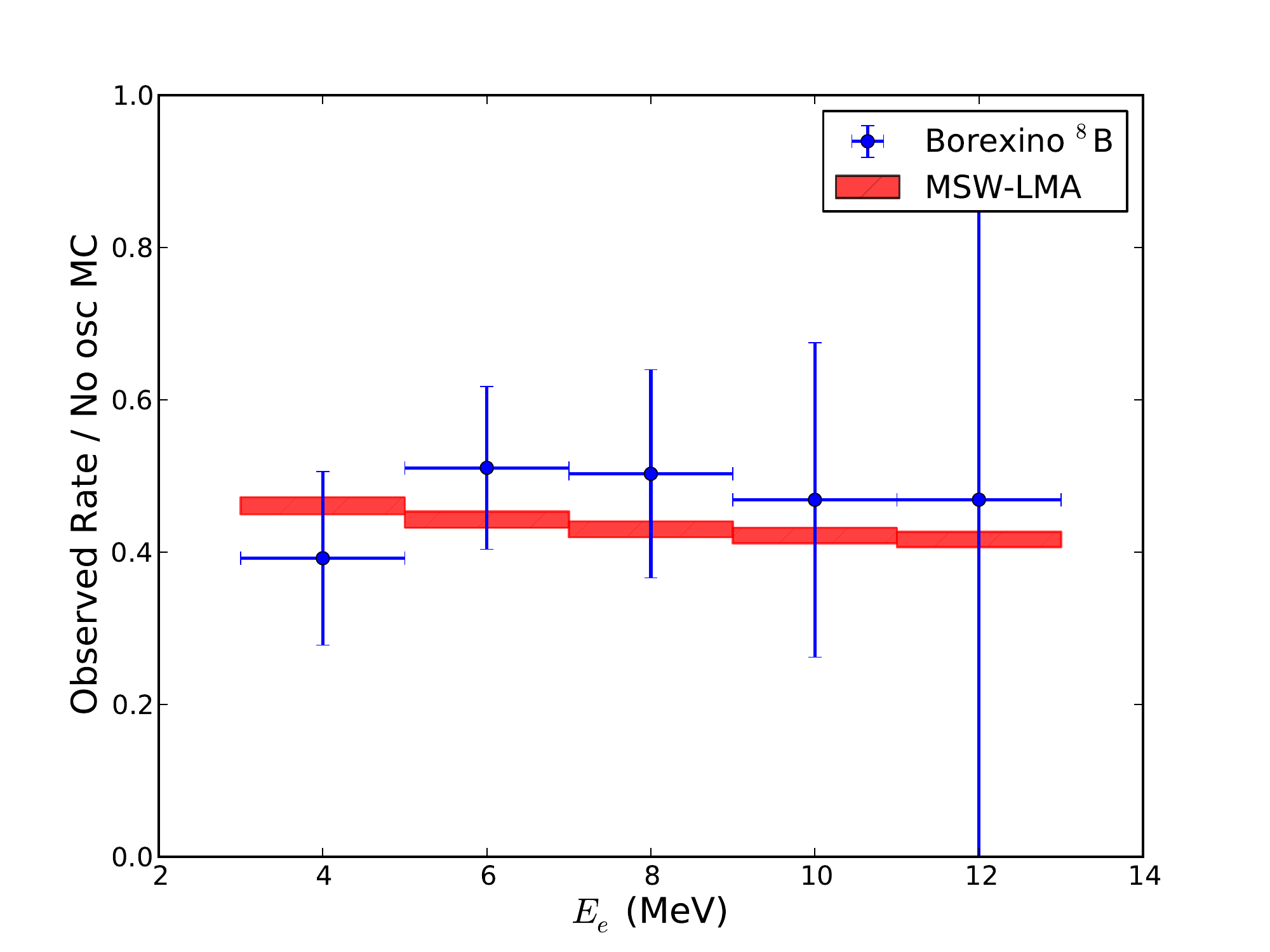}
  \caption{ \label{FIG:borexinodata}
    (Color online) Borexino event rate binned in measured electron energy with
    each bin scaled by Monte Carlo predictions assuming GS98SF2 fluxes, versus
    the same ratio for the expected rates assuming our best fit LMA parameters
    and fluxes.  Error bars on the data points represent statistical
    uncertainties only. The best fit oscillation prediction band width
    represents the uncertainty on the $^8$B flux.
  }
\end{figure}

\begin{figure}
  \includegraphics[width=0.5\textwidth]{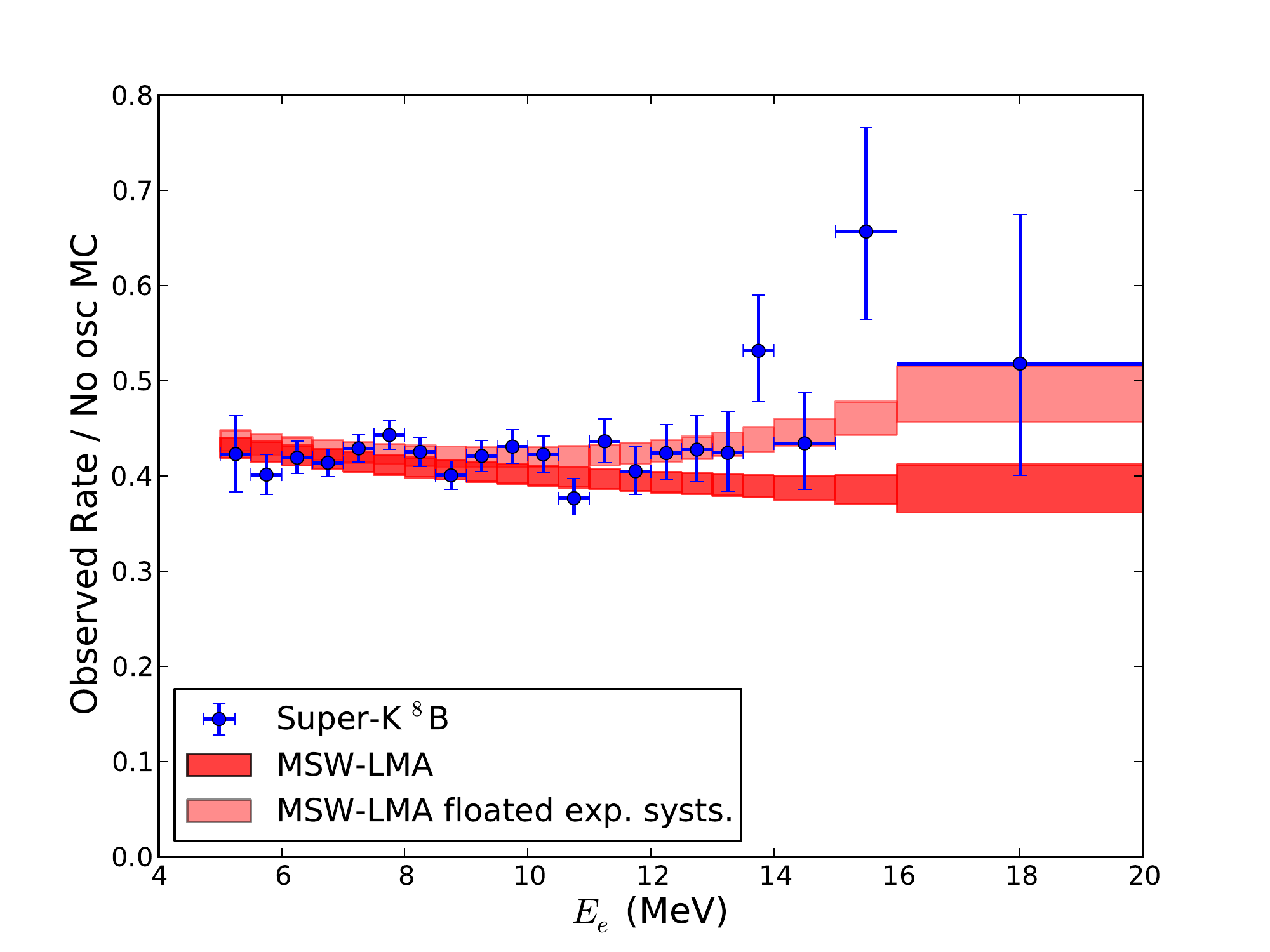}
  \caption{ \label{FIG:superk1data}
    (Color online) S-K I event rates binned in measured electron energy with
    each bin scaled by Monte Carlo predictions assuming GS98SF2 fluxes, versus
    the same ratio for the expected rates assuming our combined best fit LMA
    parameters and fluxes. Error bars on the data points represent statistical
    and energy uncorrelated systematic uncertainties combined in quadrature.
    The two bands show the effect of the correlated systematic uncertainties:
    for the dark band, detector response parameters have been fixed at their
    reported values, while for the light they have been floated in the fit. The
    best fit oscillation prediction band width represents the uncertainty on
    the $^8$B flux.
  }
\end{figure}

\begin{figure}
  \includegraphics[width=0.5\textwidth]{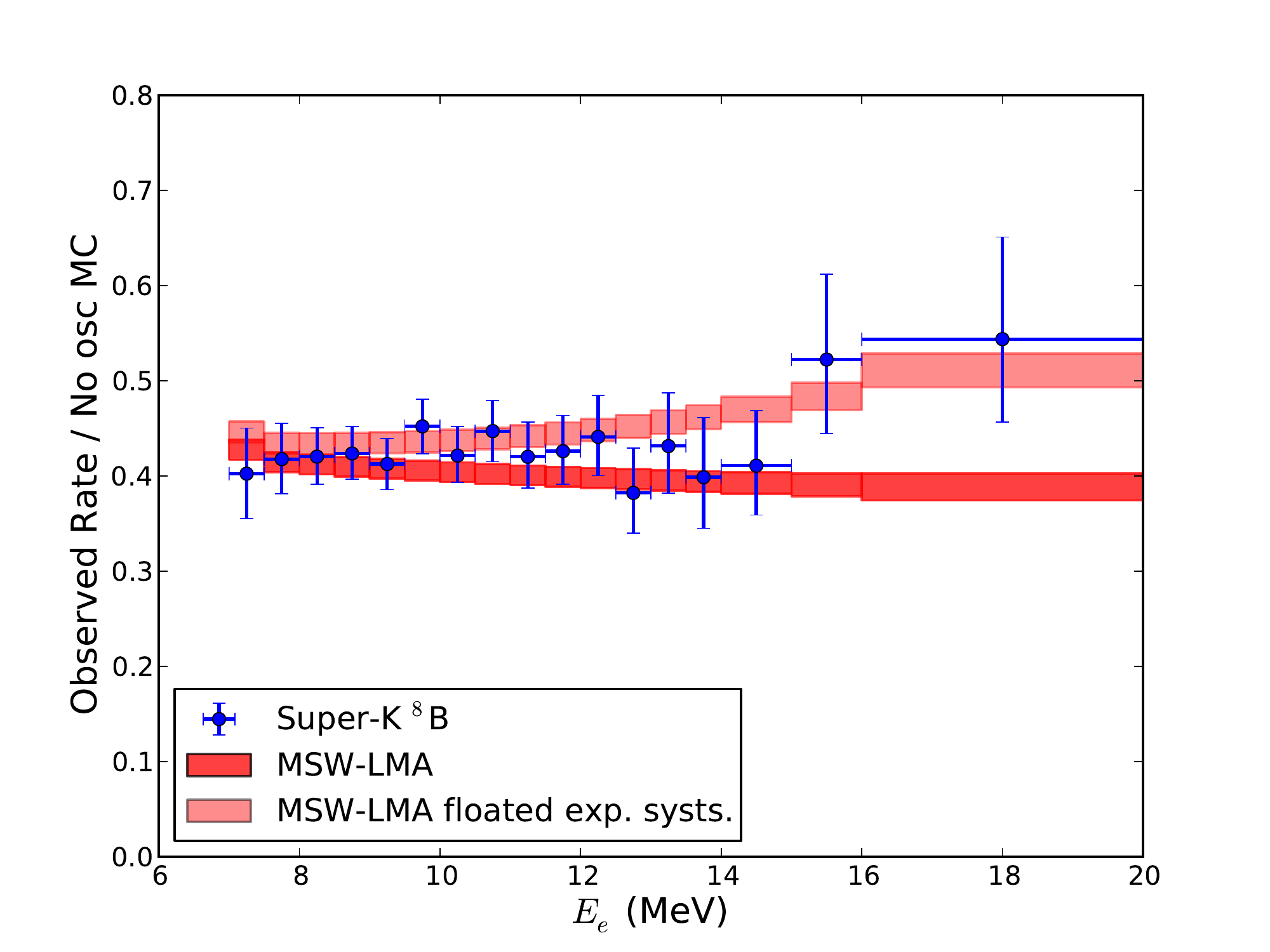}
  \caption{ \label{FIG:superk2data}
    (Color online) S-K II event rates binned in measured electron energy with
    each bin scaled by Monte Carlo predictions assuming GS98SF2 fluxes, versus
    the same ratio for the expected rates assuming our combined best fit LMA
    parameters and fluxes. Error bars on the data points represent statistical
    and energy uncorrelated systematic uncertainties combined in quadrature.
    The two bands show the effect of the correlated systematic uncertainties:
    for the dark band, detector response parameters have been fixed at their
    reported values, while for the light they have been floated in the fit. The
    best fit oscillation prediction band width represents the uncertainty on
    the $^8$B flux.
  }
\end{figure}

\begin{figure}
  \includegraphics[width=0.5\textwidth]{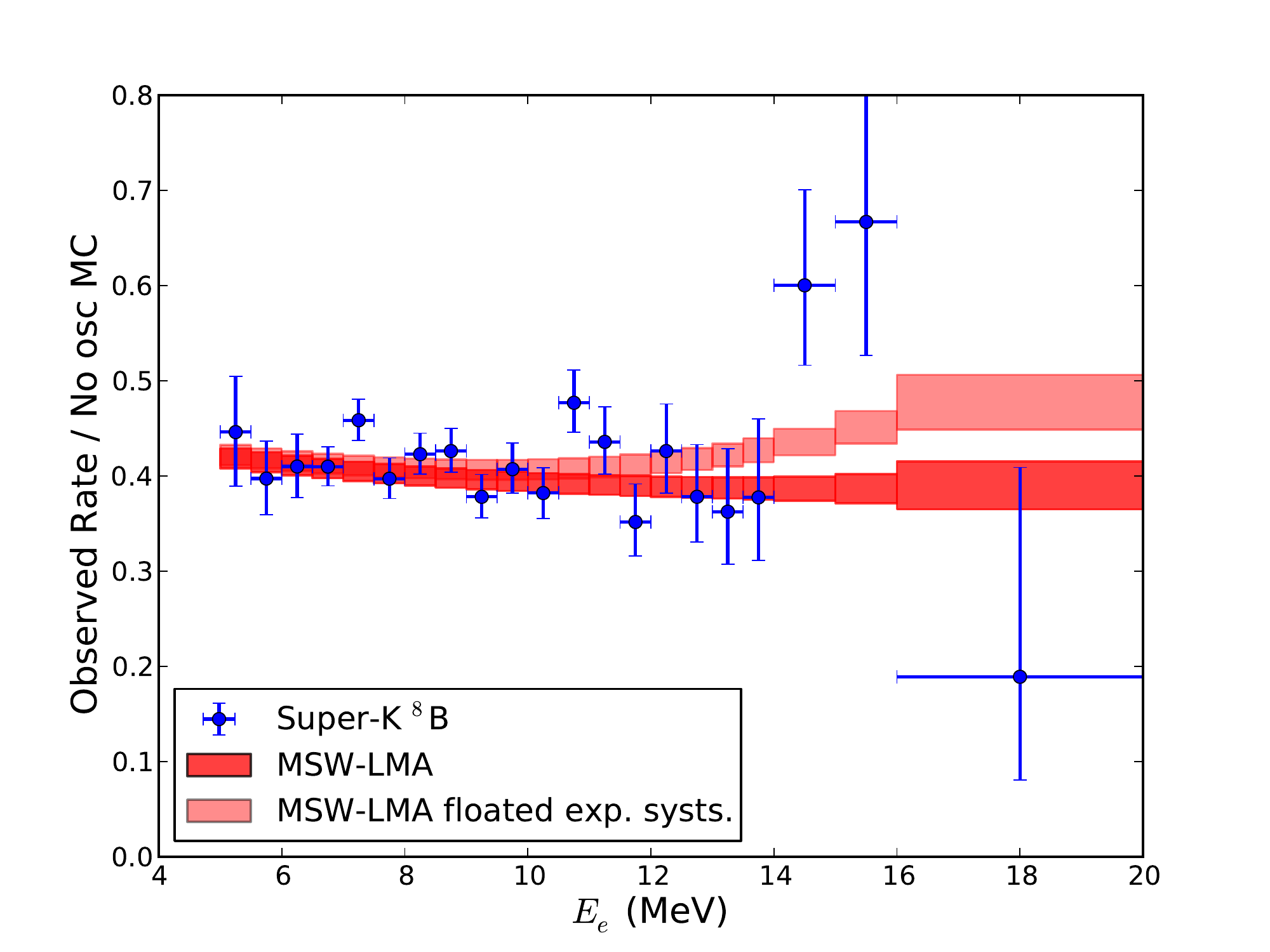}
  \caption{ \label{FIG:superk3data}
    (Color online) S-K III event rates binned in measured electron energy with
    each bin scaled by Monte Carlo predictions assuming GS98SF2 fluxes, versus
    the same ratio for the expected rates assuming our combined best fit LMA
    parameters and fluxes. Error bars on the data points represent statistical
    and energy uncorrelated systematic uncertainties combined in quadrature.
    The two bands show the effect of the correlated systematic uncertainties:
    for the dark band, detector response parameters have been fixed at their
    reported values, while for the light they have been floated in the fit. The
    best fit oscillation prediction band width represents the uncertainty on
    the $^8$B flux.
  }
\end{figure}

We find the best fit point for standard MSW-LMA at $\Delta
m^2_{21}=7.462\e{-5}$ eV$^2, \sin^2\theta_{12}=0.301,
\sin^2\theta_{13}=0.0242$, with a $^8$B flux of $5.31\e{6}$ cm$^{-2}$s$^{-1}$.
The fit compared to the data sets of SNO, Borexino, and S-K is shown in Figs.
\ref{FIG:mswb8vdata}-\ref{FIG:superk3data}.  Although in general for the
analyses in this paper we marginalize over S-K's systematic uncertainties, it
is important to note how they affect the goodness of the fit. To show this
effect, we plot the observed rate in S-K against the predicted rate calculated
from our best fit mixing parameters in two ways: first fixing the energy scale,
energy resolution, and efficiency to the values reported by S-K, and second
using values for these parameters obtained by floating them in our fit. In both
cases the width of the band does not include any of the systematic
uncertainties associated with these parameters since they are energy dependent
and so cannot be captured in a single plot. We find the best fit with the
energy scale at $+1.1\sigma$, the energy resolution at $-1.0\sigma$ and the
overall efficiency at $+0.6\sigma$. The efficiency systematic uncertainty
increases the average predicted ratio while the other two each bend up the high
energy end of the spectrum. In other words, while the LMA prediction appears to
be a poor fit to the high-energy region of the S-K data, the allowed variation
from S-K's systematic uncertainties can explain the difference if they are
moved roughly 1$\sigma$ from their central values.  Better constraints on the
S-K detector response parameters might therefore lead to a more significant
disagreement with the LMA model.

\subsection{Non-Standard Forward Scattering}

We formulate our results for this section to be comparable to Palazzo
\cite{palazzo}, so $\epsilon^e_{\alpha\beta} = \epsilon^u_{\alpha\beta}= 0$.
For a more general case to first order $n_f/n_e$ can be considered constant in
the Sun, thus any combination of $\epsilon^{e,u,d}$'s would just be a scaling
of our results.

\begin{figure}
  \includegraphics[width=0.5\textwidth]{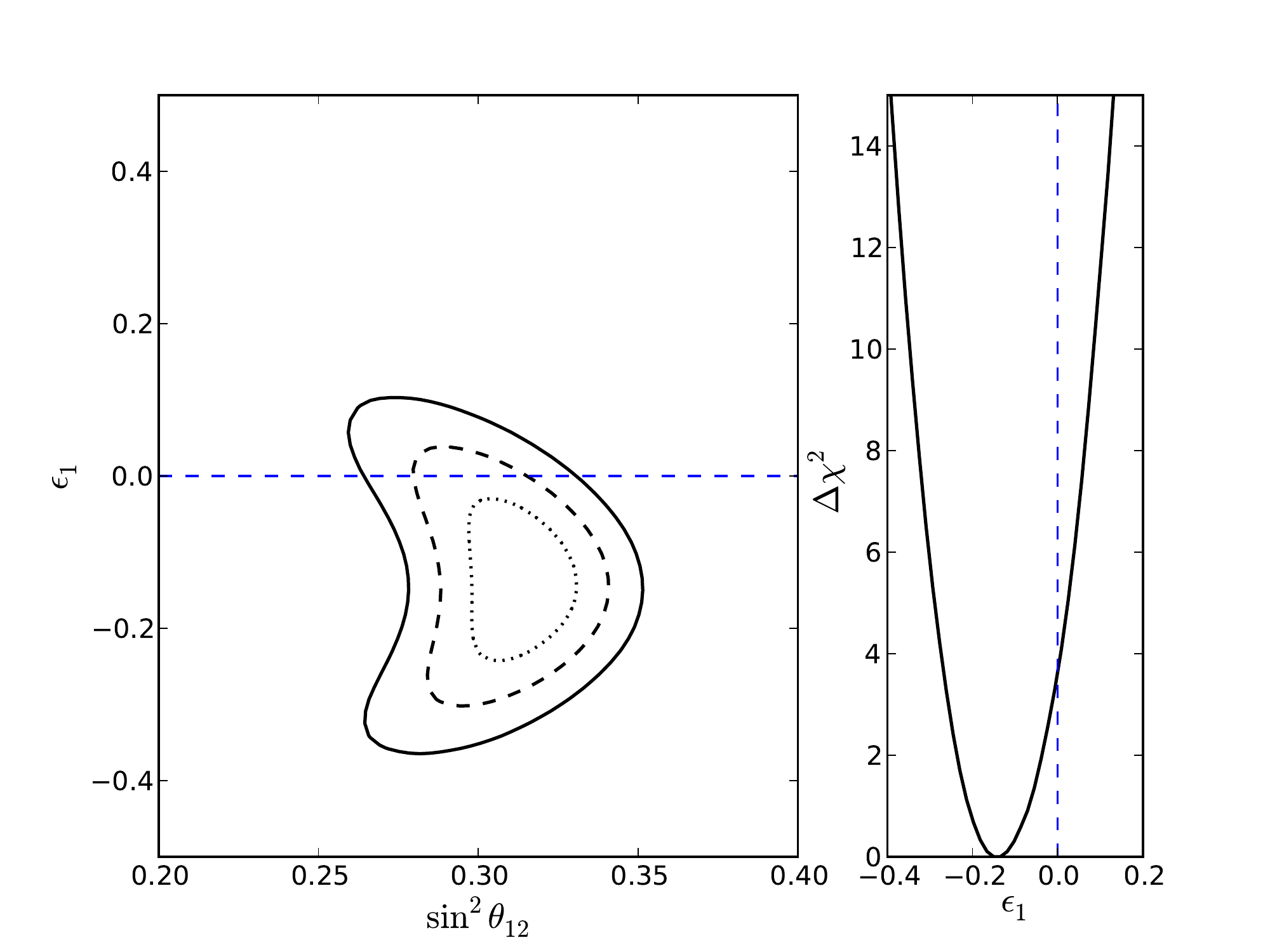}
  \caption{ \label{FIG:twoflavornsi} 
    Left: Two flavor contours with $\epsilon_2=0$ and real $\epsilon_1$.
    Contours are shown for 68\%, 95\%, and 99.73\% confidence levels for 2
    d.o.f., where the $\chi^2$ has been minimized with respect to all
    undisplayed parameters. Right: $\Delta \chi^2$ as a function of
    $\epsilon_1$.
  }
\end{figure}

\begin{figure}
  \includegraphics[width=0.5\textwidth]{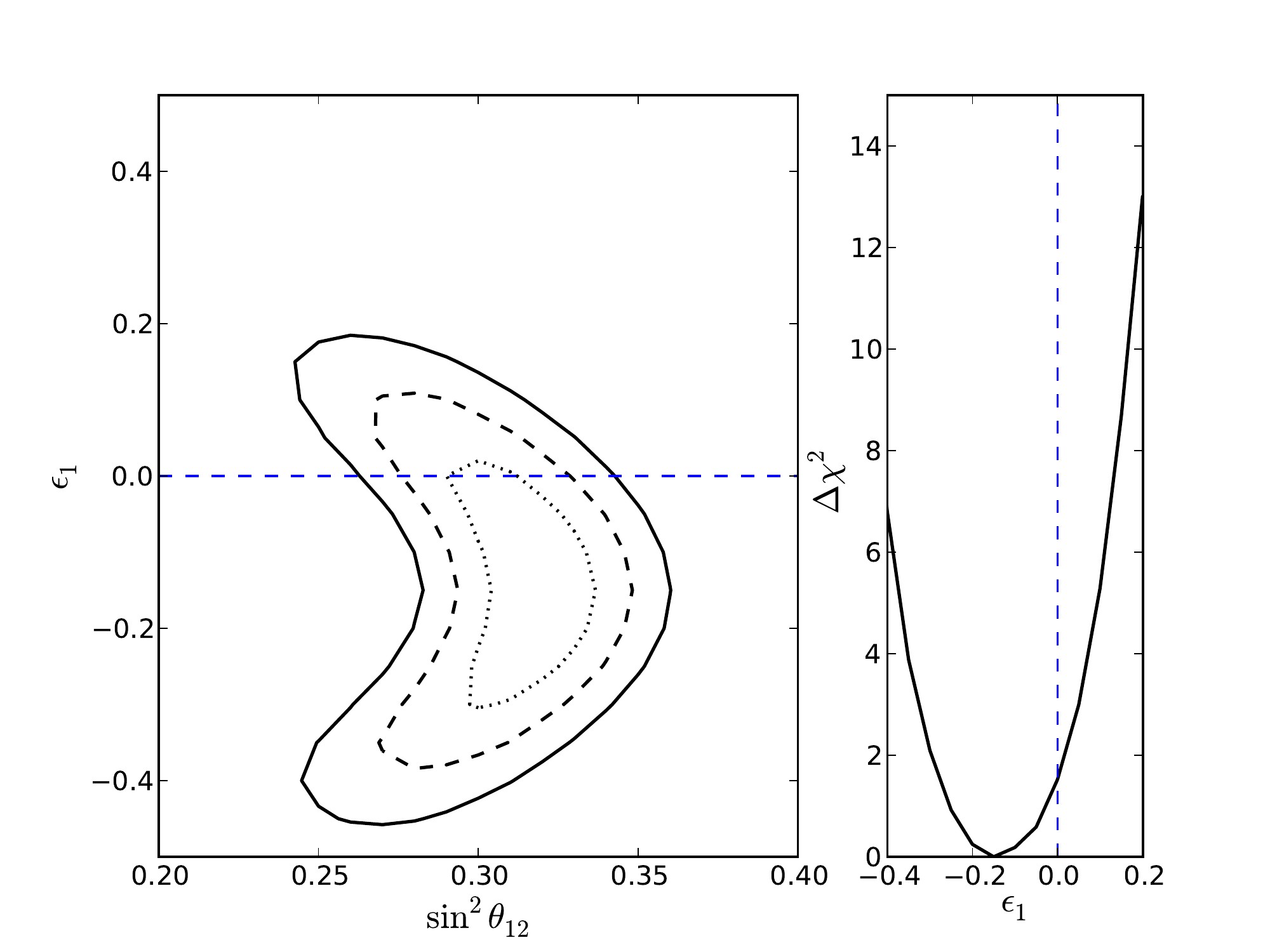}
  \caption{ \label{FIG:threeflavornsit12} 
    Left: Three flavor contours including constraints from RENO and Daya Bay.
    Contours are shown for 68\%, 95\%, and 99.73\% confidence levels for 2
    d.o.f., where the $\chi^2$ has been minimized with respect to all
    undisplayed parameters. Right: $\Delta \chi^2$ as a function of
    $\epsilon_1$.
  }
\end{figure}

\begin{figure}
  \includegraphics[width=0.5\textwidth]{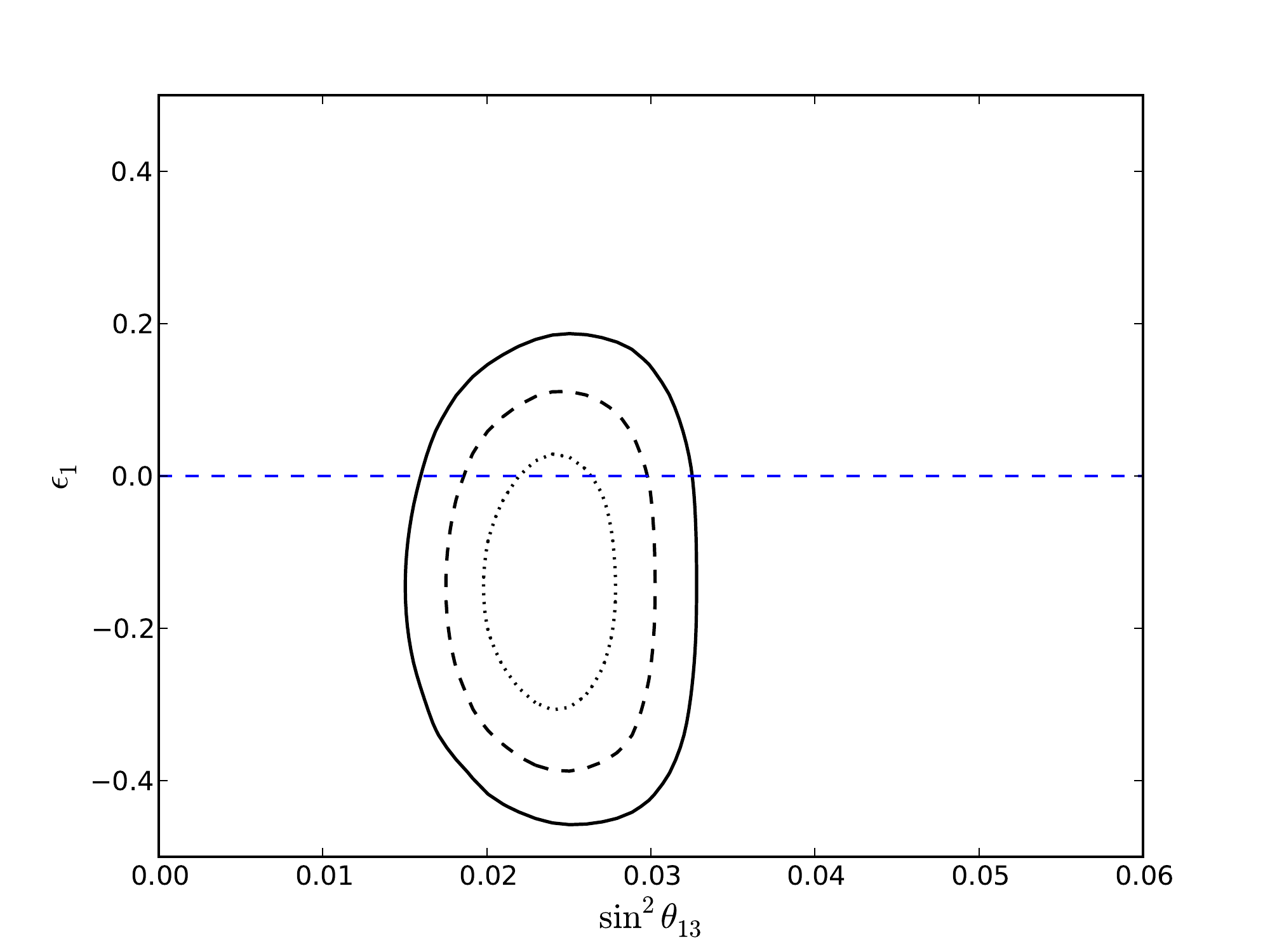}
  \caption{ \label{FIG:threeflavornsit13} 
    Three flavor contours including constraints from RENO and Daya Bay for
    $\epsilon_1$ and $\theta_{13}$. Contours are shown for 68\%, 95\%, and
    99.73\% confidence levels for 2 d.o.f., where the $\chi^2$ has been
    minimized with respect to all undisplayed parameters.
  }
\end{figure}

First we consider only real $\epsilon_1$ with $\epsilon_2=0$. Including the
most up-to-date solar results and the most recent KamLAND results as a
constraint, letting $\theta_{12}$ and $\Delta m^{2}_{12}$ float and fixing
$\theta_{13}=0$, we get a best fit of $\epsilon_1 = -0.137^{+0.070}_{-0.071}$,
shown in Fig.  \ref{FIG:twoflavornsi}, which well matches results from Palazzo.
After letting $\theta_{13}$ float and adding in the constraint from RENO and
Daya Bay, the significance becomes smaller, with a best fit value of
$\epsilon_1 = -0.145^{+0.118}_{-0.109}$, shown in Figs.
\ref{FIG:threeflavornsit12} and \ref{FIG:threeflavornsit13}. The best fit
survival probability compared to MSW-LMA and to data considered in this
analysis is shown in the appendix in Fig.  \ref{FIG:nsibestfit}.

\begin{figure}
  \includegraphics[width=0.5\textwidth]{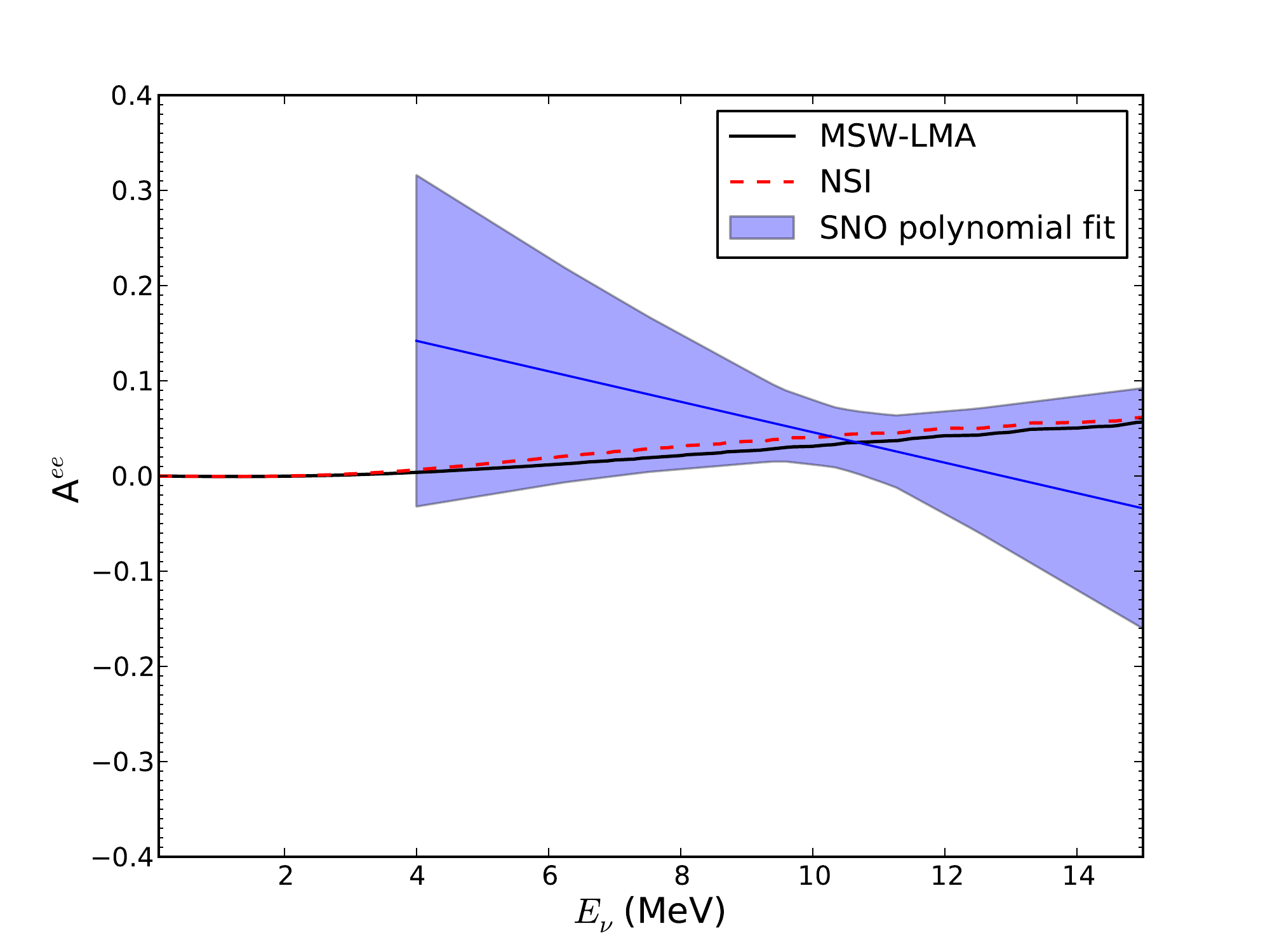}
  \caption{ \label{FIG:asymmpolysno} 
    (Color online) Day-Night asymmetry from SNO results compared to best fit
    MSW-LMA and NSI.  The band represents the RMS spread at any given energy,
    i.e., not including energy correlations.
  }
\end{figure}

\begin{figure}
  \includegraphics[width=0.5\textwidth]{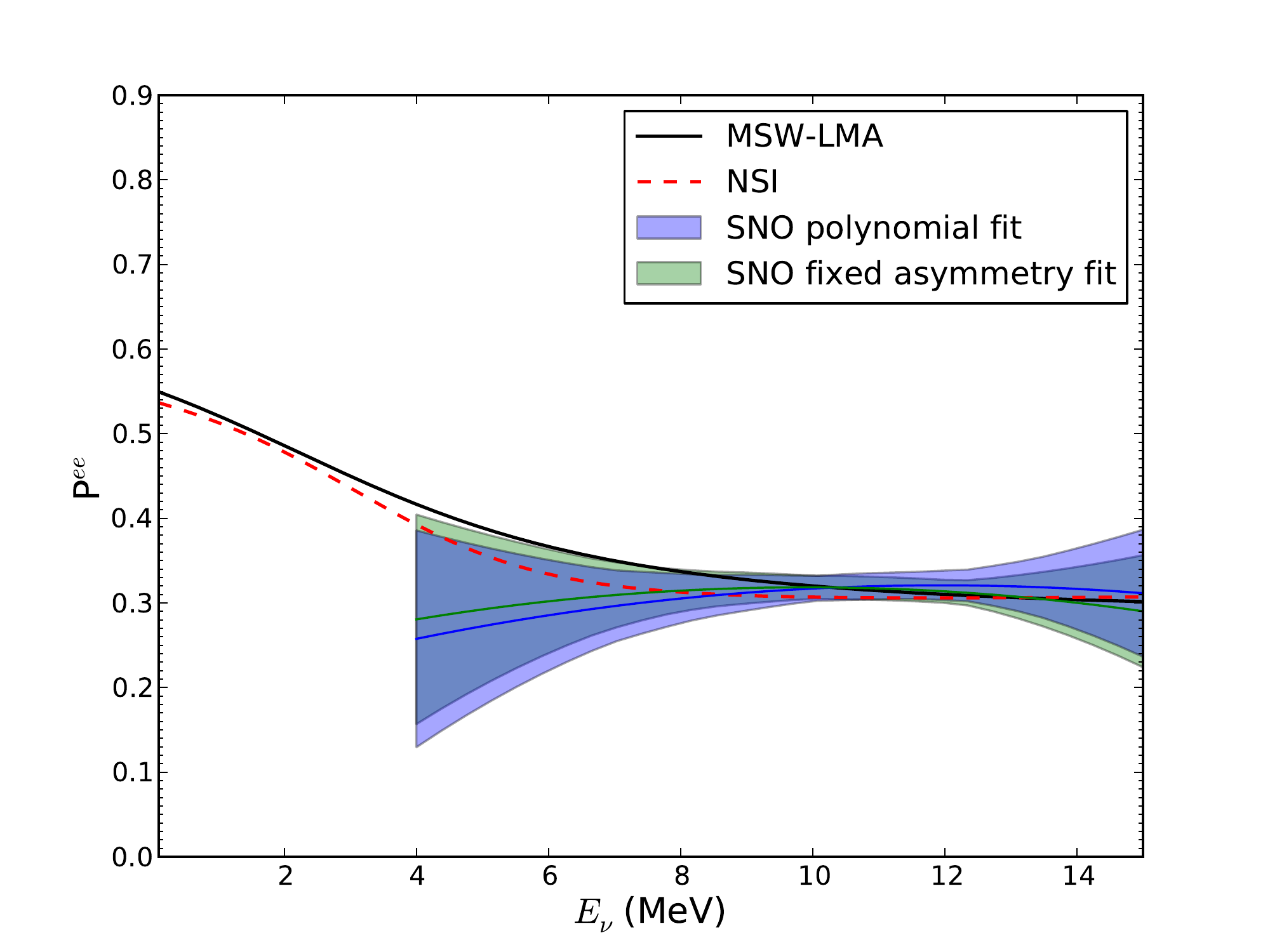}
  \caption{ \label{FIG:fixedasymmsno} 
    (Color) Day survival probability for SNO. The blue band shows the RMS
    spread from the best fit, and the green band shows the spread after the
    Day-Night asymmetry is fixed to the MSW-LMA prediction.
  }
\end{figure}

These results seem to allow a vacuum to matter transition in the survival
probability at higher energies than the SNO data suggests. It is important to
consider the fit to the day night asymmetry, shown in Fig.
\ref{FIG:asymmpolysno} for SNO. The NSI does not have a large effect on the
asymmetry, and so for both models the best fit does not fit the data well. The
correlations between the asymmetry and the day survival probability translate
this poor fit to an even broader allowed upturn, further limiting the
significance of any flatness in the data. We show this effect by fitting the
MSW-LMA predicted day-night asymmetry to Eq. \ref{EQ:snopolydaynight} and then
recalculating what the RMS spread in the day night survival probability would
be after fixing $a_0$ and $a_1$ given the correlation matrix, as shown in
\ref{FIG:fixedasymmsno}.

In addition, since these plots scale the absolute rates to get survival
probabilities, they hide the relationship between the survival probability and
the absolute flux. Both of these effects can be seen more clearly in the
correlation matrix for SNO's polynomial survival probability fit, Table VIII in
Ref. \cite{sno3phase}. The baseline level of the survival probability $c_0$ is
strongly anticorrelated with the absolute flux $\Phi_{B}$, and the slope of the
survival probability $c_1$ is anticorrelated with the slope of the day night
asymmetry $a_1$. 

\begin{figure}
  \includegraphics[width=0.5\textwidth]{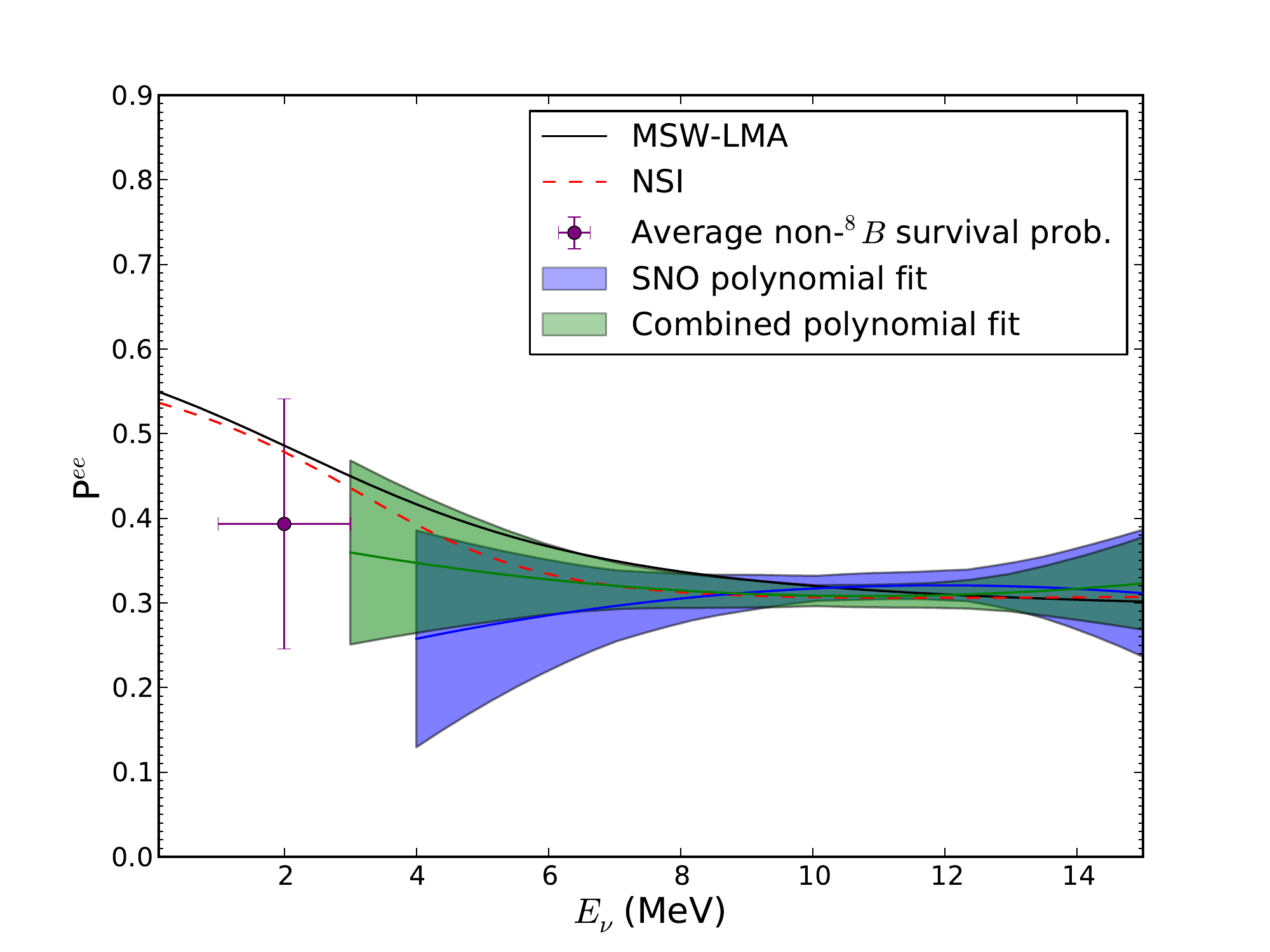}
  \caption{ \label{FIG:combinedpoly} 
    (Color) Polynomial fit to SNO, Super-Kamiokande, Borexino $^8$B data and
    Homestake's results. The band represents the RMS spread at any given
    energy, i.e., not including energy correlations.
  }
\end{figure}

To better visualize why the full fit does not have a better constraint, we
applied the polynomial survival probability fit as used for the SNO data to the
combination of the SNO, S-K, Borexino, and Homestake results.  This represents
a fit to the survival probability independent of any physics model, where the
polynomial forms in Eqs. \ref{EQ:snopolyday} and \ref{EQ:snopolydaynight} are
used to impose an energy correlation under the model independent assumption
that there is no small scale structure to the survival probability. Since the
Homestake results could contain a significant fraction of non-$^8$B events, one
additional term for the average non-$^8$B survival probability is added to the
fit, where non-$^8$B fluxes were fixed at SSM values. The results of the fit
are given in Tables \ref{TAB:combofit} and \ref{TAB:combocor}, and the best fit
and RMS spread is shown in Fig. \ref{FIG:combinedpoly}. The majority of the
change from the SNO-only band is driven by the S-K results, where the
high-energy end and the $^8B$ flux is pulled upward. Although their data looks
flat in detected energy, when projected back into incident neutrino energy it
becomes consistent with an LMA-like transition, as suggested in Figs.
\ref{FIG:superk1data}, \ref{FIG:superk2data}, and \ref{FIG:superk3data}. The
band of the RMS spread shows the significance to which we can say anything
about the the shape of the survival probability at low energies, and we can see
that the band covers the MSW-LMA prediction but at the same time allows for a
perfectly flat or even downward bending survival probability. Note that this
combined polynomial fit does not impact any of the results in this paper since
we are only using it to visualize the survival probability and do not actually
use it in our likelihood fits.

\begin{table}
\begin{tabular}{lll}
      \hline
      \hline
       & Best Fit & Fit Error\\
      \hline
      $\Phi_B$ & 5.403 & 0.195 \\
      $c_0$ & 0.309 & 0.015 \\
      $c_1$ & -0.0014 & 0.0055 \\
      $c_2$ & 0.008 & 0.0022 \\
      $a_0$ & 0.047 & 0.020 \\
      $a_1$ & 0.000 & 0.018 \\
      $P_{\text{non-}^8B}$ & 0.393 & 0.148 \\
      \hline
      \hline
    \end{tabular}
    \caption{\label{TAB:combofit}
      Results for polynomial fit for the survival probability and day-night
      asymmetry fit to the data of SNO, S-K, Borexino, and Homestake.
  }
\end{table}

\begin{table}
\begin{tabular}{llllllll}
      \hline
      \hline
      & $\Phi_B$ & $c_0$ & $c_1$ & $c_2$ & $a_0$ & $a_1$ & $P_{\text{non-}^8B}$\\
      \hline
      $\Phi_B$ & 1.000 & -0.793 & 0.215 & -0.152 & -0.027 & 0.016 & 0.045 \\
      $c_0$ & -0.793 & 1.000 & -0.289 & -0.279 & -0.204 & -0.009 & -0.074 \\
      $c_1$ & 0.215 & -0.289 & 1.000 & -0.010 & 0.042 & -0.587 & 0.023 \\
      $c_2$ & -0.152 & -0.279 & -0.010 & 1.000 & -0.032 & -0.004 & -0.073 \\
      $a_0$ & -0.027 & -0.204 & 0.042 & -0.032 & 1.000 & -0.073 & 0.014 \\
      $a_1$ & 0.016 & -0.009 & -0587 & -0.004 & -0.073 & 1.000 & 0.005 \\
      $P_{\text{non-}^8B}$ & 0.045 & -0.074 & 0.023 & -0.073 & 0.014 & 0.005 & 1.000 \\
      \hline
      \hline
    \end{tabular}
    \caption{\label{TAB:combocor}
      Correlation matrix from the polynomial fit for the survival probability
      and day-night asymmetry fit to the data of SNO, S-K, Borexino, and
      Homestake.
    }
\end{table}

\begin{figure}
  \includegraphics[width=0.5\textwidth]{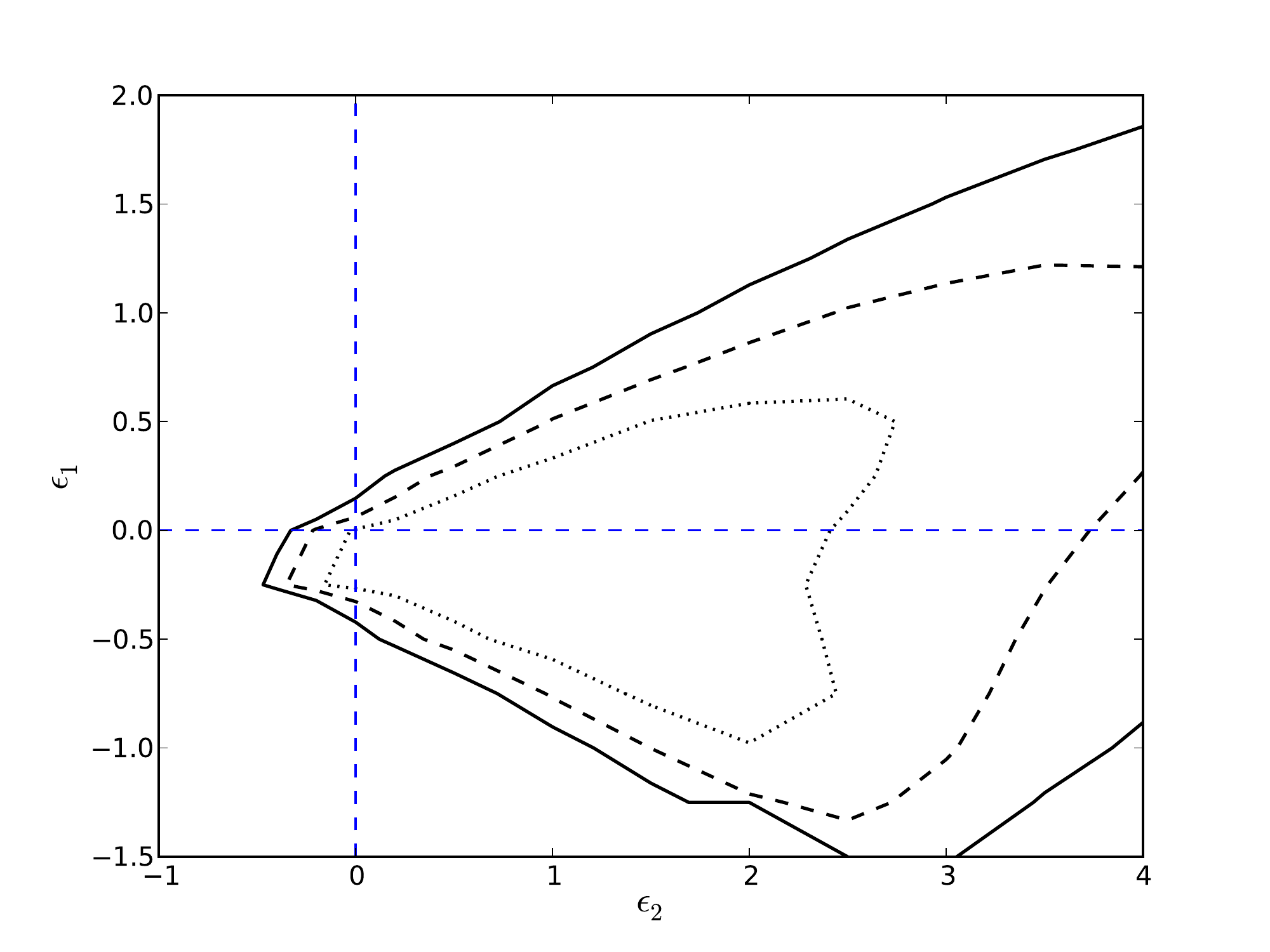}
  \caption{ \label{FIG:twoparamnsi} 
    Results for NSI fit with real $\epsilon_1$. Contours are shown for 68\%,
    95\%, and 99.73\% confidence levels (2 d.o.f.), where the $\chi^2$ has been
    minimized with respect to all undisplayed parameters.
  }
\end{figure}

We also consider the case of complex $\epsilon_1$. Here the best fit is found
at $\epsilon_1=-0.146 + 0.031i$. The fit results are shown in Fig.
\ref{FIG:complexnsi} and the best fit survival probability in the appendix in
Fig.  \ref{FIG:complexnsibestfit}. For both $\epsilon_1$ and $\epsilon_2$
nonzero, we find the best fit point at $\epsilon_1=0.014, \epsilon_2=0.683$.
The fit contours are shown in Fig. \ref{FIG:twoparamnsi}, and the best fit
survival probability is shown in Fig.  \ref{FIG:twoparamnsibestfit}. In both
cases the additional free parameter allows a slightly better fit, but the
standard MSW-LMA is within the 68\% confidence interval for two degrees of
freedom. Once both $\epsilon_1$ and $\epsilon_2$ are allowed to be nonzero,
there is no further improvement in the fit if we again let $\epsilon_1$ be
complex.

For all of these scenarios, the best fit values for the non-standard parameters
$\epsilon_1$ and $\epsilon_2$ are well within the current experimental bounds.
At the same time, they represent relatively substantial effects, considering
that at $\epsilon_{\alpha\beta} = 1$ the non-standard interaction has the same
strength as the MSW potential, as shown in Eq. \ref{EQ:threeflavornsi}.

\begin{figure}
  \includegraphics[width=0.5\textwidth]{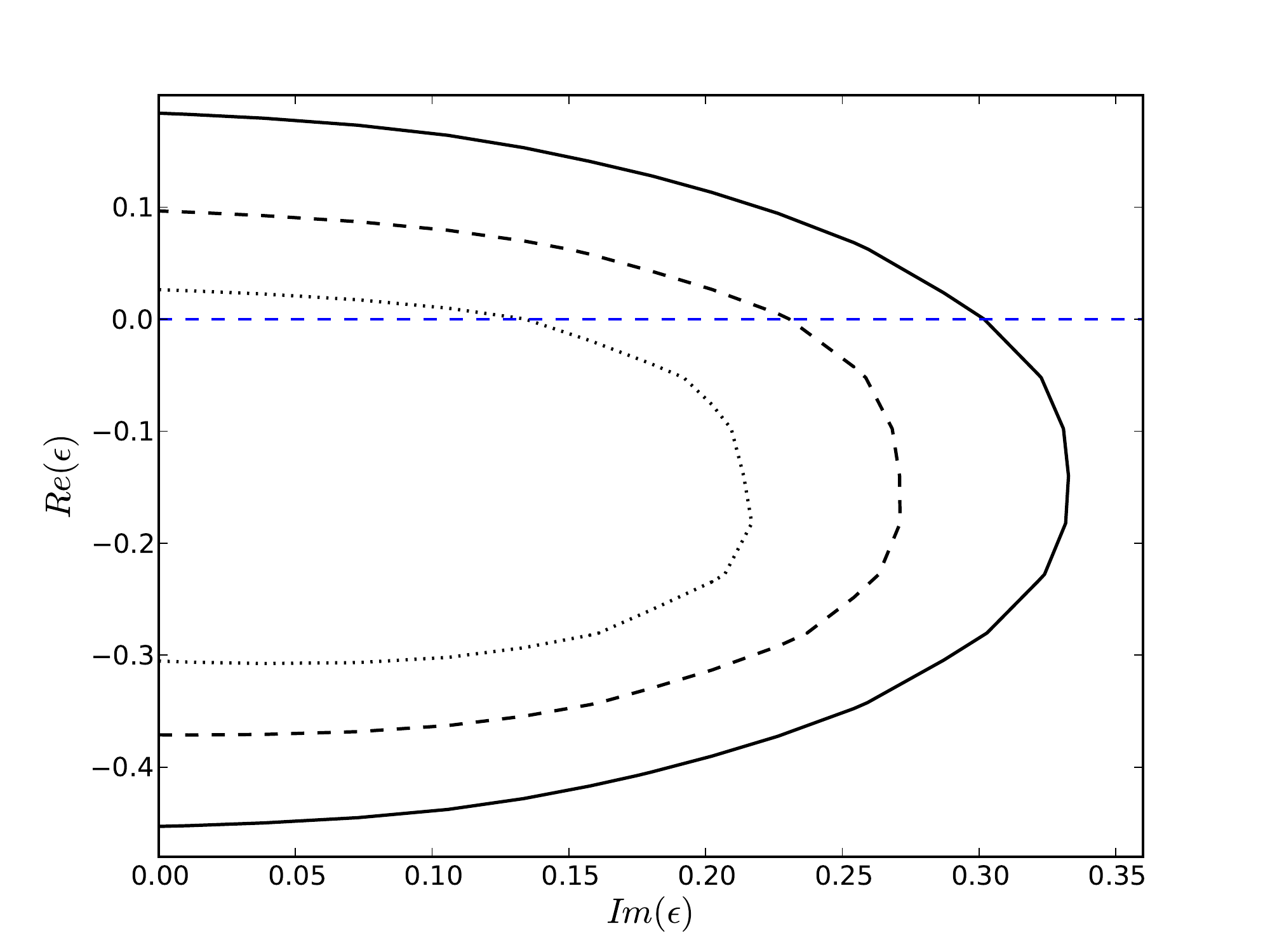}
  \caption{ \label{FIG:complexnsi} 
    Results for NSI fit with $\epsilon_2=0$ but complex $\epsilon_1$. Contours
    are shown for 68\%, 95\%, and 99.73\% confidence levels (2 d.o.f.), where
    the $\chi^2$ has been minimized with respect to all undisplayed parameters.
  }
\end{figure}

\subsection{Mass Varying Neutrinos}

\subsubsection{Neutrino Density Effects}

\begin{figure}
  \includegraphics[width=0.5\textwidth]{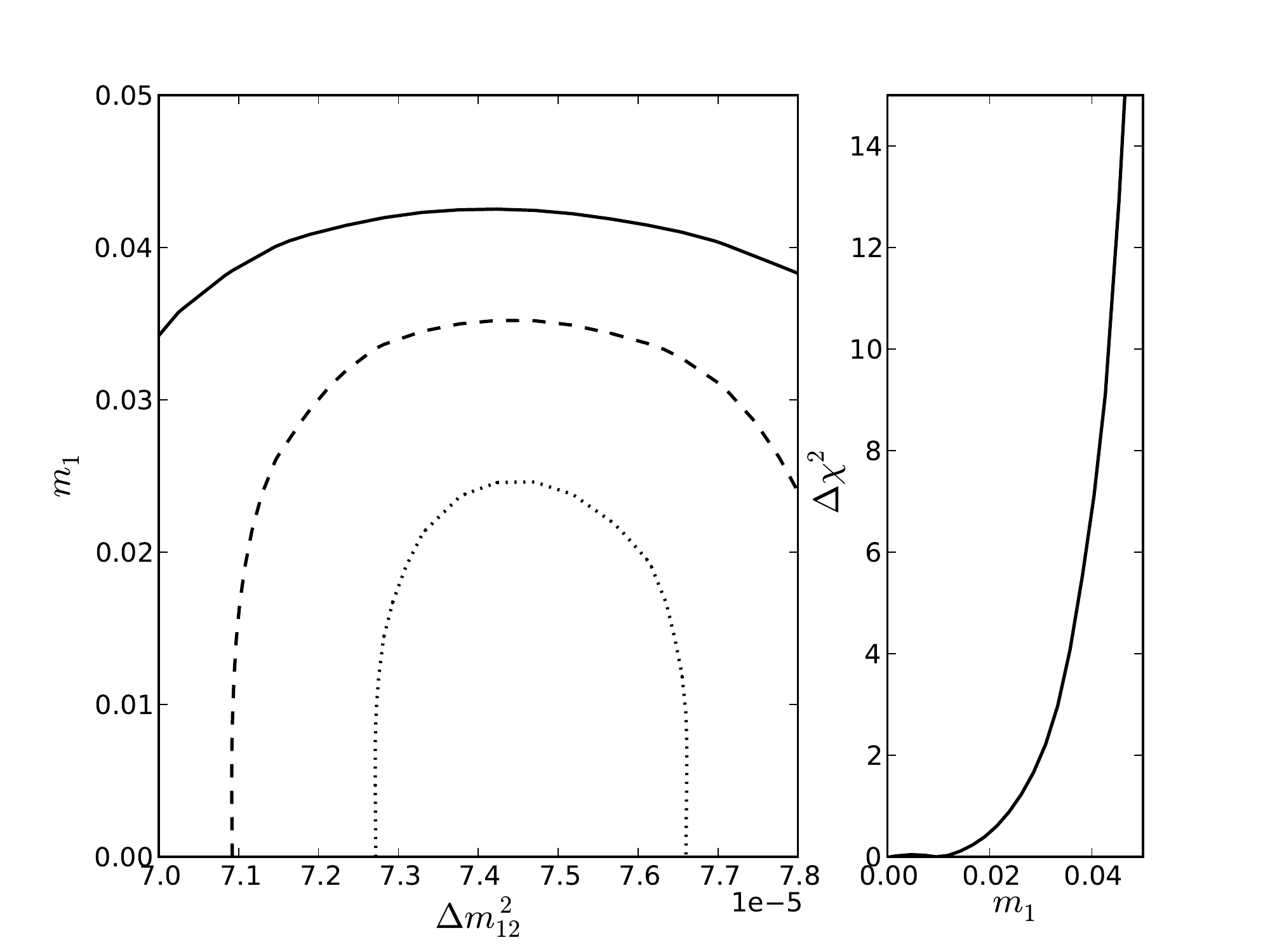}
  \caption{ \label{FIG:mavannu} 
    Results for MaVaN model with neutrino mass coupled to neutrino density.
    Left: Contours are shown for 68\%, 95\%, and 99.73\% confidence levels for
    2 d.o.f., where the $\chi^2$ has been minimized with respect to all
    undisplayed parameters. Right: $\Delta \chi^2$ as a function of $m_{1,0}$.
  }
\end{figure}

After fitting for $m_{1,0}$ letting all mixing parameters float, we found that
the best fit point was at $m_{1,0}=0$, where this model's predictions become
identical to MSW-LMA. Our fit results, as shown in Fig. \ref{FIG:mavannu}, give
a 90\% confidence level upper limit on the neutrino mass scale of $m_{1,0} <
0.033$eV within this model. Our results do not agree with the previous limit in
\cite{mavancirelli}, who found a limit an order of magnitude smaller. We cannot
explain the difference, although they use older data sets for each experiment.
For the inverted hierarchy we expect $m_{1,0} \gae \sqrt{\Delta m^2_{atm}} \sim
0.05$ eV, so within the context of this model, the inverted hierarchy would be
rejected.

\subsubsection{Fermion Density Effects}

\begin{figure}
  \includegraphics[width=0.5\textwidth]{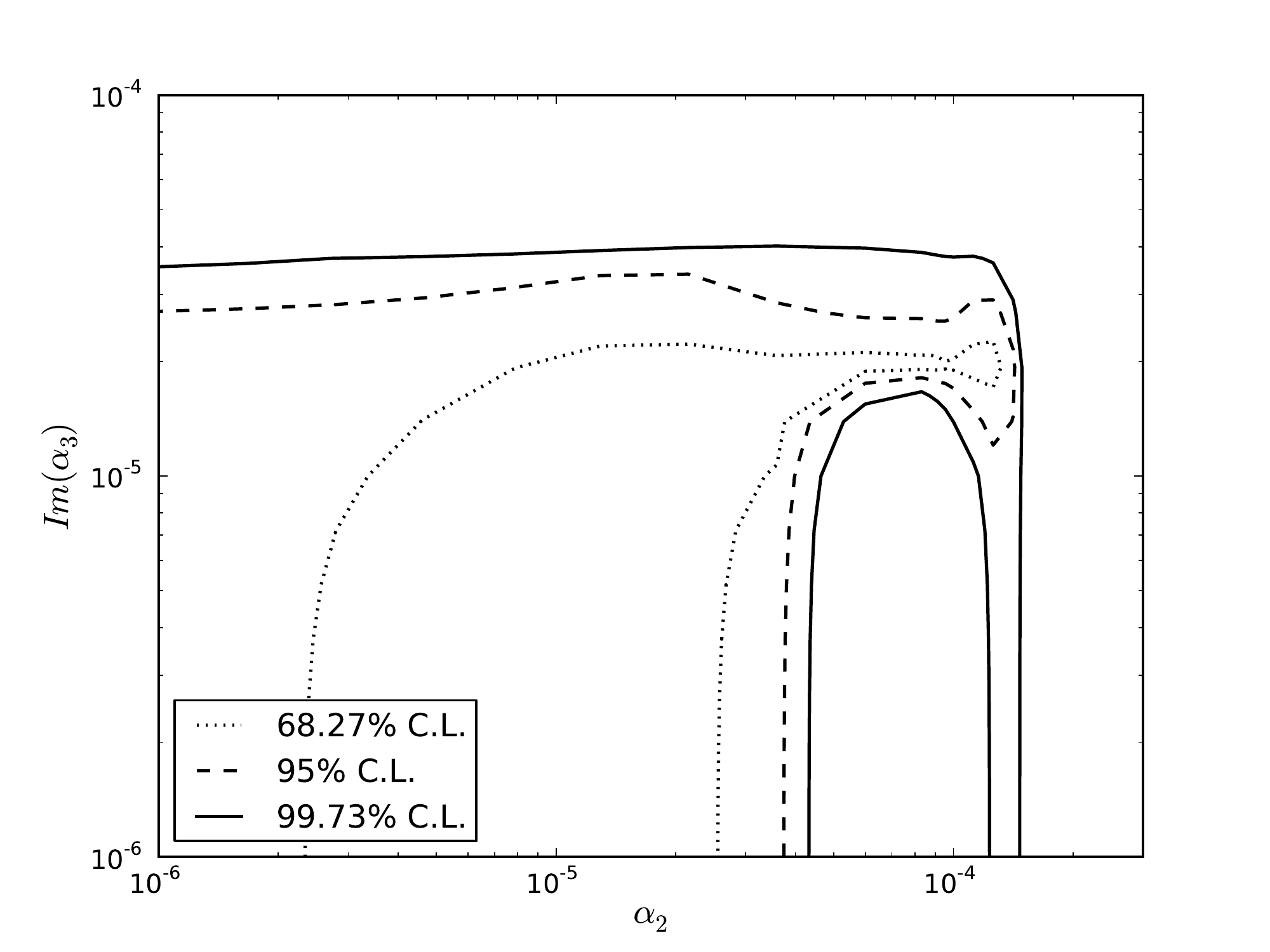}
  \caption{ \label{FIG:mavanfermi} 
    Results for MaVaN model with neutrino mass coupled to fermion density with
    $\alpha_2 > 0$ and $\alpha_3^2 < 0$. Contours are shown for 68\%, 95\%, and
    99.73\% confidence levels (2 d.o.f.), where the $\chi^2$ has been minimized
    with respect to all undisplayed parameters.
  }
\end{figure}

For simplification we let $m_{1,0} = \alpha_{1} = 0$, so we fit for
$\alpha_2,Re[\alpha_3],Im[\alpha_3]$. Results for $\alpha_2 > 0$, $\alpha_3^2 <
0$ are shown in Fig. \ref{FIG:mavanfermi}. In this case our best fit is at
$\alpha_2=5.95\e{-5}$, $\alpha_3=i1.97\e{-5}$, shown in the appendix in Fig.
\ref{FIG:mavan2bestfit}, although the $2\sigma$ contour includes the origin.
Note that although the $^8$B survival probability in Fig.
\ref{FIG:mavan2bestfit} seems to be far from the Borexino $pep$ point, in this
scenario the $pep$ survival probability is actually significantly different
than $^8$B's at the same energy, making it more consistent with the data than
it would appear.  Minimizing over all other variables gives the bounds at
90\% confidence for 1 d.o.f. of

\begin{eqnarray}
  1.6\e{-6} \leq &\alpha_2/\text{eV}& \leq 1.3\e{-4}, \\
                  &|\alpha_3|/\text{eV}& \leq 2.48\e{-5} \text{ for } \alpha_3^2 > 0, \\
                  &|\alpha_3|/\text{eV}& \leq 2.29\e{-5} \text{ for } \alpha_3^2 < 0. 
\end{eqnarray}

Then from Eq. \ref{EQ:mavanfermi}, we can use our limits on the parameters to
get a combined limit on the couplings of $|\lambda^{ij}\lambda^{N}|/m_\phi^2
\leq 2.8\e{-14}$eV$^{-2}$ \cite{mavangonzalez}. 

\subsection{Long-Range Leptonic Forces}

For the scalar long-range leptonic force, we find that after again fixing
$m_{1,0} = 0$, the best fit is at $k_S = 6.73\e{-45}$, $\lambda = 1.56
R_{\odot}$.  Since $\lambda = 1/m_S$, this point represents a force mediated by
a scalar particle with mass $m_S = 9.1\e{-17}$eV and a coupling strength
$g_0=2.91\e{-22}$. The best fit survival probability is shown in the appendix
in Fig.  \ref{FIG:lrscalarbestfit}.  Like the MaVaN case, the $pep$ survival
probability is higher than $^8$B's at the same energy. For the long-range
vector force, we find the best fit at $k_V = 3.26\e{-54}, \lambda = 16.97
R_{\odot}$, shown in Fig.  \ref{FIG:lrvectorbestfit}. For the tensor long-range
force, there is no improvement of the fit to the data and the best fit remains
at MSW-LMA.  

\begin{figure}
  \includegraphics[width=0.5\textwidth]{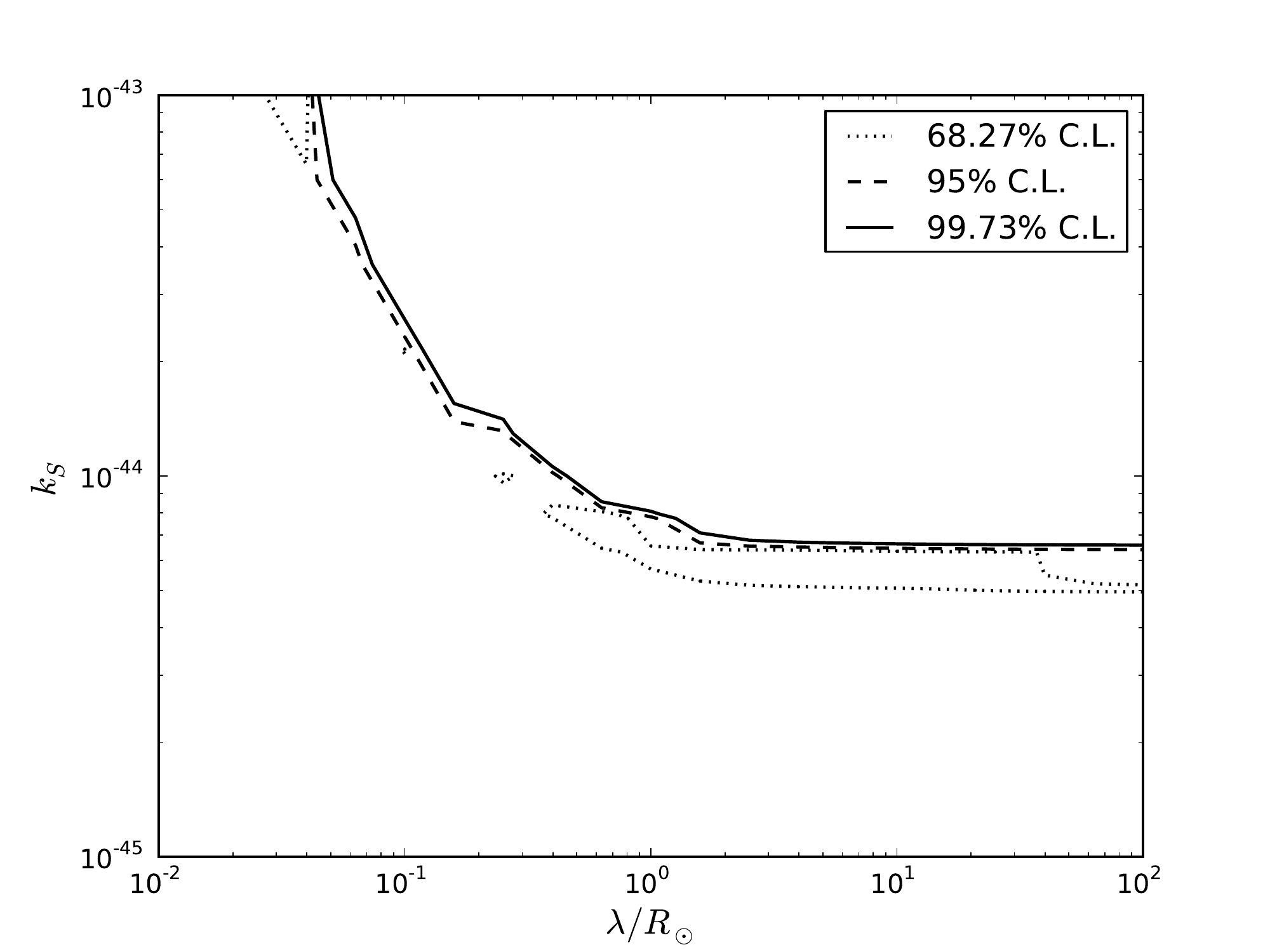}
  \caption{ \label{FIG:lrscalar}
    Results for MaVaN model with a scalar long-range force and $m_{1,0} = 0$.
    Contours are shown for 68\%, 95\%, and 99.73\% confidence levels (2
    d.o.f.), where the $\chi^2$ has been minimized with respect to all
    undisplayed parameters.
  }
\end{figure}

In all three cases, standard MSW-LMA is within the $1\sigma$ contour, but the
constraint on the coupling strength gets stronger as $\lambda$ increases.  The
contours for the scalar case are shown in Fig. \ref{FIG:lrscalar}. At $\lambda
= \infty$, we can set upper limits on the coupling strengths at 90\% confidence
level for 1 d.o.f. of

\begin{eqnarray}
  k_S(e) &\leq& 6.31\e{-45}\text{ with }m_1=0\text{eV}, \\
  k_V(e) &\leq& 1.23\e{-53}, \\
  k_T(e) &\leq& 1.31\e{-61}\text{eV}^{-1}.
\end{eqnarray}

\subsection{Non-Standard Solar Model}

We found that using the low metallicity (AGSS09SF2) solar model's flux
constraints and solar distributions did not give noticeably different results,
and in general worsened the fits for any model.

\begin{figure}
  \includegraphics[width=0.5\textwidth]{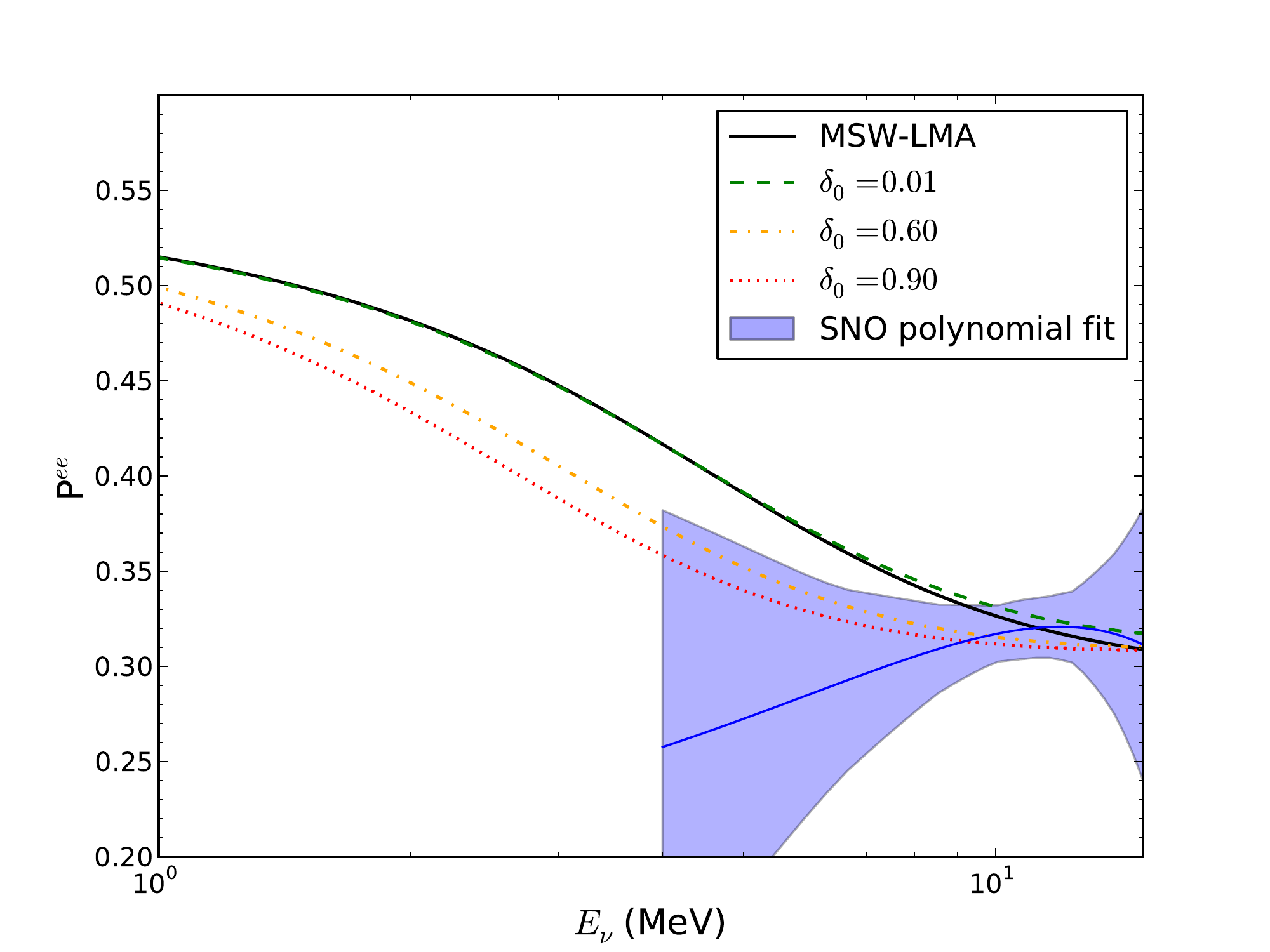}
  \caption{ \label{FIG:solardensityrange}
    (Color online) Survival probability for MSW-LMA with various fractional
    increases $\delta_0$ of the solar core density compared to the SNO results.
  }
\end{figure}

As described in Section \ref{SEC:nssolarmodel}, we looked at the effect of
changing the density of the solar core to see whether we are susceptible to
mistaking a small difference in the expected solar model for a non-standard
interaction. Fig.  \ref{FIG:solardensityrange} shows the survival probability
with the core density increased by various amounts. It is clear that within the
range suggested by helioseismological measurements of about 1\%, the change in
the $^8$B upturn is not large enough to mimic any of the non-standard models.
Fitting for the central density while keeping the rest of the fit the same, we
find that the improvement in the fit for a change of up to 1\% is marginal, and
we don't reach a minimum until an implausible increase in the solar core
density of around 90\%. Since any change in the central density would change
the core temperature and thus also the expected fluxes, we fit again allowing
the density to float and replacing the flux constraints from the solar model
with an overall luminosity constraint and a constraint on the $pp$ to $pep$
ratio. Here the best fit is found at an increase of 57\%, with a $\Delta
\chi^2$ of $-4.6$, although not changing the density and just removing the flux
constraints already gives a $\Delta \chi^2$ of -3.5.

\section{Conclusions}

We have compared the predictions of survival probabilities for several models
of neutrino non-standard interactions compared to standard MSW-LMA oscillations
using results from solar experiments constrained by terrestrial measurements of
the mixing parameters. The results of the fits are summarized in Table
\ref{TAB:summary}.

Although several of these models allow for a better fit to the data and suggest
an explanation for the flatness of the $^8$B survival probability, we have
shown that with the current available data on solar neutrino interactions,
there is no model that has demonstrated to be better than MSW-LMA with greater
than $2 \sigma$ significance. We have found that the low significance is in
part due to the known, large value of $\theta_{13}$, but also because of the
as-yet large systematic uncertainties and covariances in the experimental data
sets. The critical transition region thus remains largely unexplored. 

We have also examined whether small changes to the solar density profile could
lead to a change in the transition region that could mimic the effects of new
physics. The results of our simple model show that in fact this is not the
case.  The matter/vacuum transition region is therefore a good place to look
for small effects of non-standard models.

Our best fit survival probabilities show that because most of our non-standard
model effects have a solar radial or density dependence, the effect is lessened
in the $pep$ or $pp$ production regions and so it would be difficult to test
these models merely by improving the measurement of either of those signals. It
would require either a better measurement of lower energy $^8$B, especially one
with a charged-current interaction that preserves more of the spectral
information, or a new model that can more closely match the data in order for
this discrepancy to become more than a hint of something non-standard.

To fully probe this interesting region, in which the interferometry provided by
neutrino oscillations lets us look for even tiny effects of new physics, will
require new experiments or more precisely constrained systematic uncertainties.
Both the Super-Kamiokande and Borexino experiments will continue to take data
and hopefully their uncertainties will continue to improve.  The SNO+
experiment will begin taking data in the near future and it, too, will be able
to probe this region.  It is possible, however, that a measurement using a
charged-current reaction, which preserves more of the spectral information, may
be necessary to provide the needed precision to see any new physics that may
lie in this region.

  \begin{table*}
    \begin{tabular}{llccc}
      \hline
      \hline
      Model & Best Fit & $\Delta \chi^2$ & Additional D.o.F.  & C.L.\\
      \hline
      MSW-LMA & $\Delta m^2_{21}=7.462\e{-5}$ eV$^2,\sin^2\theta_{12}=0.301,$ & 0 & --- & ---\\
       & $\sin^2\theta_{13}=0.0242$ & & & \\
      MSW-LMA (AGSS09SF2) & $\Delta m^2_{21}=7.469\e{-5}$ eV$^2,\sin^2\theta_{12}=0.304,$ & 2.8 & --- & ---\\
       & $\sin^2\theta_{13}=0.0240$ & & & \\
      NSI ($\epsilon_1$ real, $\epsilon_2=0$) & $\epsilon_1=-0.145$ & -1.5 & 1 & 0.78\\
      NSI ($\epsilon_2=0$) & $\epsilon_1=-0.146+0.031i$ & -1.5 & 2 & 0.53\\
      NSI ($\epsilon_1$ real) & $\epsilon_1=0.014$,$\epsilon_2=0.683$ & -1.9 & 2 & 0.60\\
      MaVaN neutrino density dependence & $m_{1,0} < 0.033$ eV & 0 & 1 & 0.0\\
      MaVaN fermi density dependence &
      $\alpha_{2}=5.95\e{-5},\alpha_{3}=i1.97\e{-5}$ & -3.4 & 2 & 0.81\\
      Long range scalar leptonic force & $k_S = 6.73\e{-45}, \lambda = 1.56 R_\odot, m_{1,0}=0$eV & -2.9 & 3 & 0.58\\
      Long range vector leptonic force & $k_V = 3.26\e{-54}, \lambda = 16.97 R_\odot$ & -1.8 & 2 & 0.59\\ 
      Long range tensor leptonic force & $k_T < 1.3\e{-61}$eV$^{-1}$ & 0 & 2 & 0.0\\
      Non-standard solar model & $\delta_0=0.57$ & -4.6 & 1 & ---\\
      \hspace{5mm} without flux constraint & & & & \\
      \hline
      \hline
    \end{tabular}
    \caption{\label{TAB:summary}
      Comparison of survival probability fits to standard MSW-LMA. If the best
      fit remains at the MSW-LMA value for a model, a 90\% confidence level
      upper limit (1 d.o.f.) on the model's parameters is given instead.
      $\Delta \chi^2$ is the difference between the model's best fit point and
      the MSW-LMA best fit. The final column gives the largest confidence level
      at which MSW-LMA is excluded.
    }
  \end{table*}

\begin{acknowledgments}

We would like to thank the SNO collaboration for their helpful comments and for
allowing us to spot check our code against theirs, and in particular Nuno
Barros for many helpful suggestions. We also would like to thank Aldo Serenelli
for providing us with information on the solar models used in this paper,
Stefano Davini for details on Borexino's $pep$ results, and Alex Friedland and
Michael Smy for helpful and interesting conversations.  This work has been
supported by the US Department of Energy, Office of Nuclear Physics, the
University of California at Berkeley, and Lawrence Berkeley National
Laboratory.

\end{acknowledgments}

\bibliography{article}
\appendix*

\section{Survival Probability Fits}
\FloatBarrier

\begin{figure}
  \includegraphics[width=0.5\textwidth]{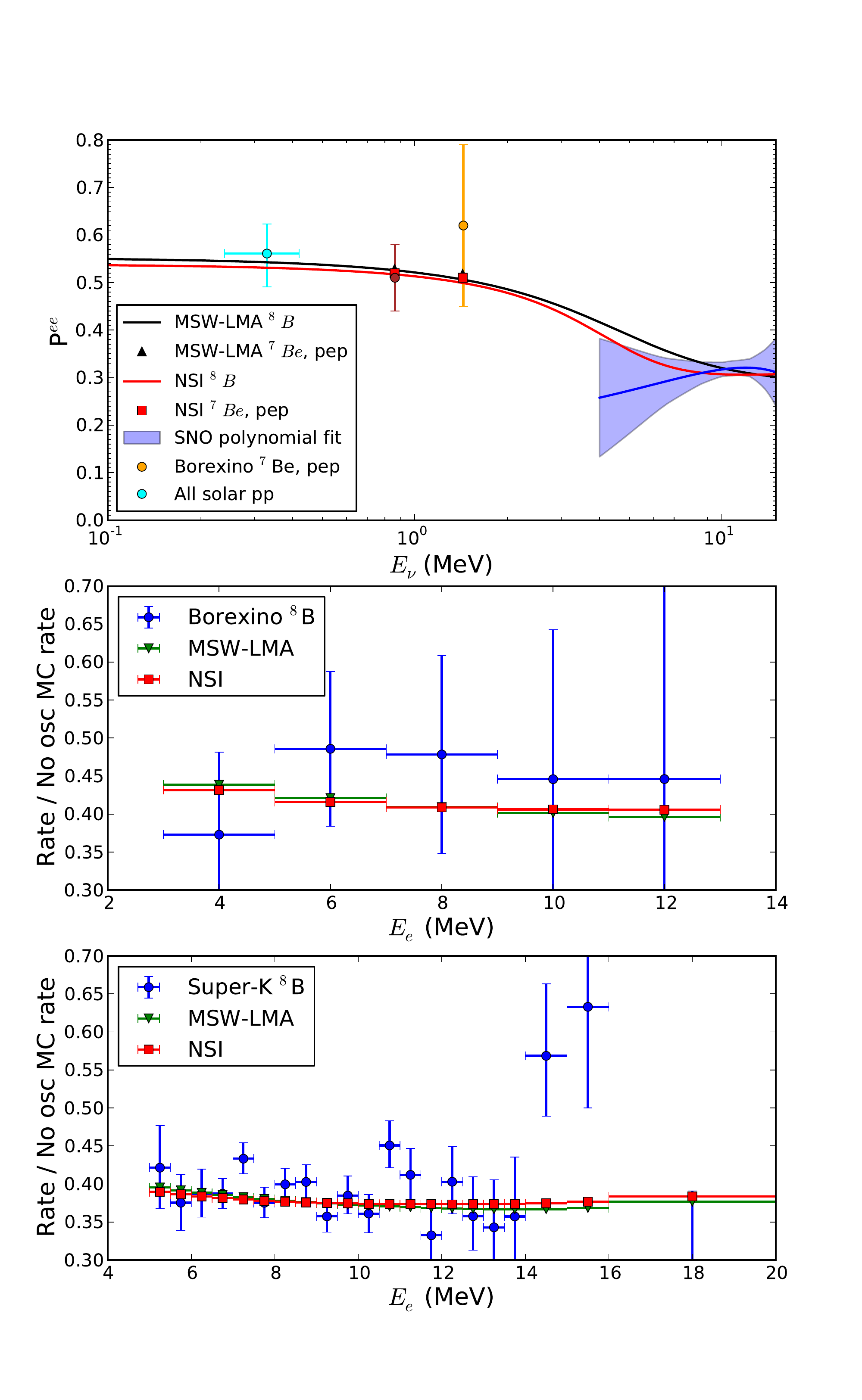}
  \caption{ \label{FIG:nsibestfit} 
    (Color online) Three flavor best fit NSI survival probability compared to
    MSW-LMA at $\epsilon_1=-0.145$, $\Delta m^2_{21}=7.481\e{-5}$eV$^2,
    \sin^2\theta_{12}=0.320, \sin^2\theta_{13}=0.0238$.  The top plot shows the
    survival probability as a function of incident neutrino energy. The middle
    shows the best fit's predicted event rate in Borexino for each of
    Borexino's measured electron energy bins scaled by the GS98SF2 flux
    no-oscillation prediction compared to Borexino's data, and the bottom shows
    the same for S-K III's energy bins and data.
  }
\end{figure}

\begin{figure}
  \includegraphics[width=0.5\textwidth]{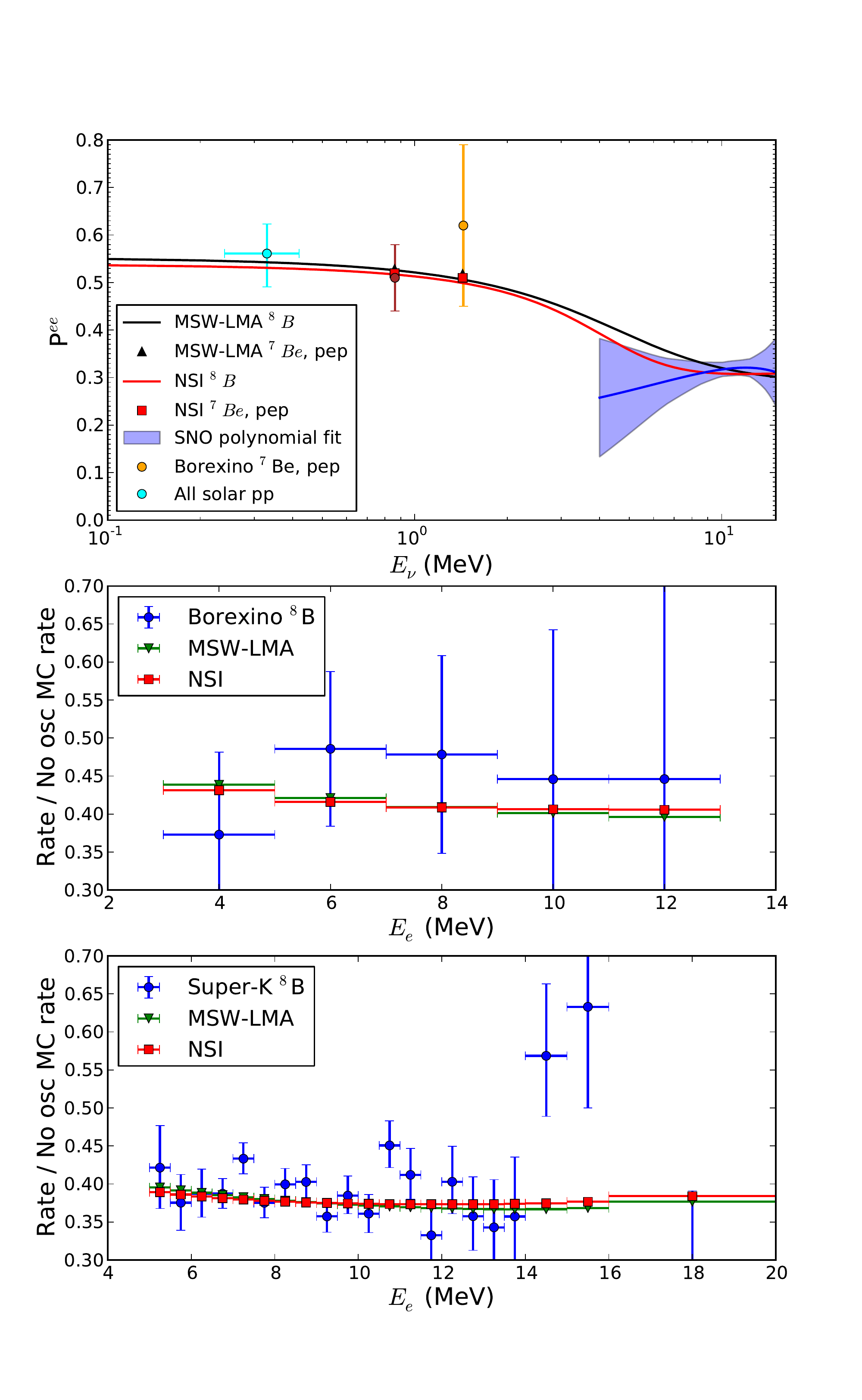}
  \caption{ \label{FIG:complexnsibestfit} 
    (Color online) Best fit for NSI fit with $\epsilon_2=0$ but complex
    $\epsilon_1$ at $\epsilon_1=-0.146+0.31i$, $\Delta
    m^2_{21}=7.472\e{-5}$eV$^2, \sin^2\theta_{12}=0.320,
    \sin^2\theta_{13}=0.0238$.  The top plot shows the survival probability as
    a function of incident neutrino energy. The middle shows the best fit's
    predicted event rate in Borexino for each of Borexino's measured electron
    energy bins scaled by the GS98SF2 flux no-oscillation prediction compared
    to Borexino's data, and the bottom shows the same for S-K III's energy bins
    and data.
  }
\end{figure}

\begin{figure}
  \includegraphics[width=0.5\textwidth]{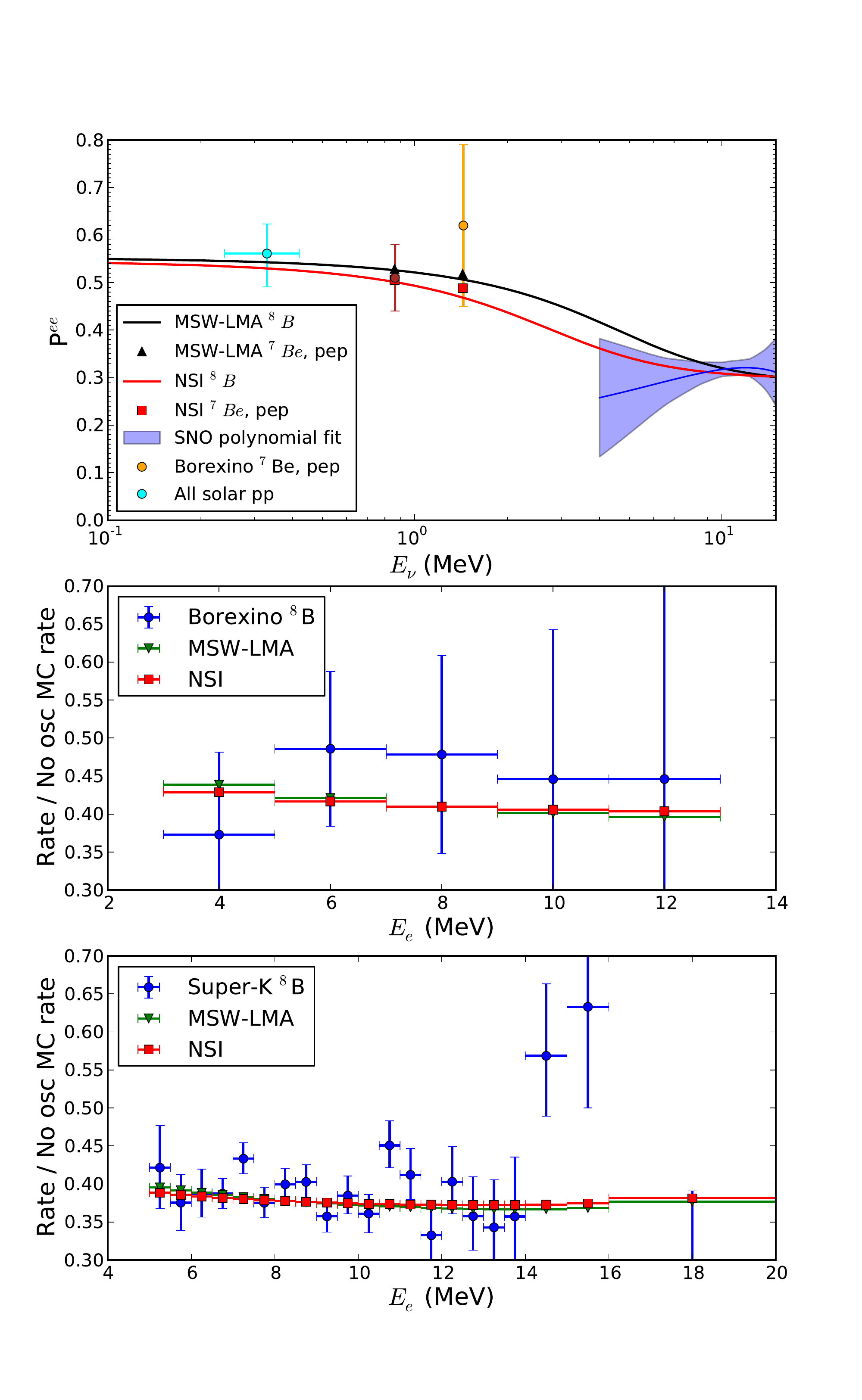}
  \caption{ \label{FIG:twoparamnsibestfit} 
    (Color online) Best fit for NSI fit with real $\epsilon_1$ at
    $\epsilon_1=0.014, \epsilon_2=0.683$, $\Delta m^2_{21}=7.487\e{-5}$eV$^2,
    \sin^2\theta_{12}=0.310, \sin^2\theta_{13}=0.0238$.  The top plot shows the
    survival probability as a function of incident neutrino energy. The middle
    shows the best fit's predicted event rate in Borexino for each of
    Borexino's measured electron energy bins scaled by the GS98SF2 flux
    no-oscillation prediction compared to Borexino's data, and the bottom shows
    the same for S-K III's energy bins and data.
  }
\end{figure}

\begin{figure}
  \includegraphics[width=0.5\textwidth]{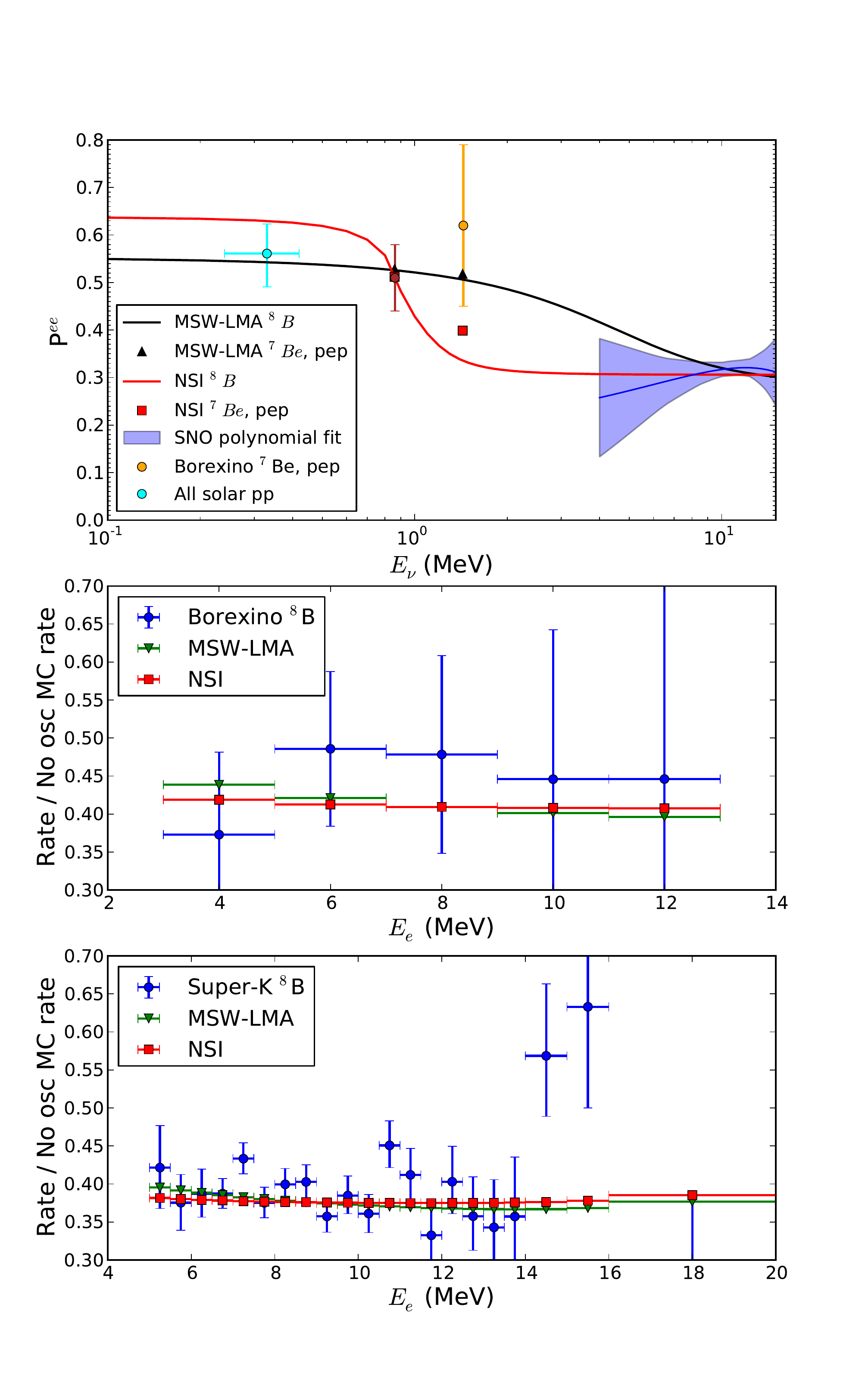}
  \caption{ \label{FIG:mavan2bestfit} 
    (Color online) Best fit for fermion density dependent MaVaN at
    $\alpha_2=5.95\e{-5}, \alpha_3=i1.97\e{-5}$, $\Delta
    m^2_{21}=7.484\e{-5}$eV$^2, \sin^2\theta_{12}=0.320,
    \sin^2\theta_{13}=0.0239$.  The top plot shows the survival probability as
    a function of incident neutrino energy. The middle shows the best fit's
    predicted event rate in Borexino for each of Borexino's measured electron
    energy bins scaled by the GS98SF2 flux no-oscillation prediction compared
    to Borexino's data, and the bottom shows the same for S-K III's energy bins
    and data.
  }
\end{figure}

\begin{figure}
  \includegraphics[width=0.5\textwidth]{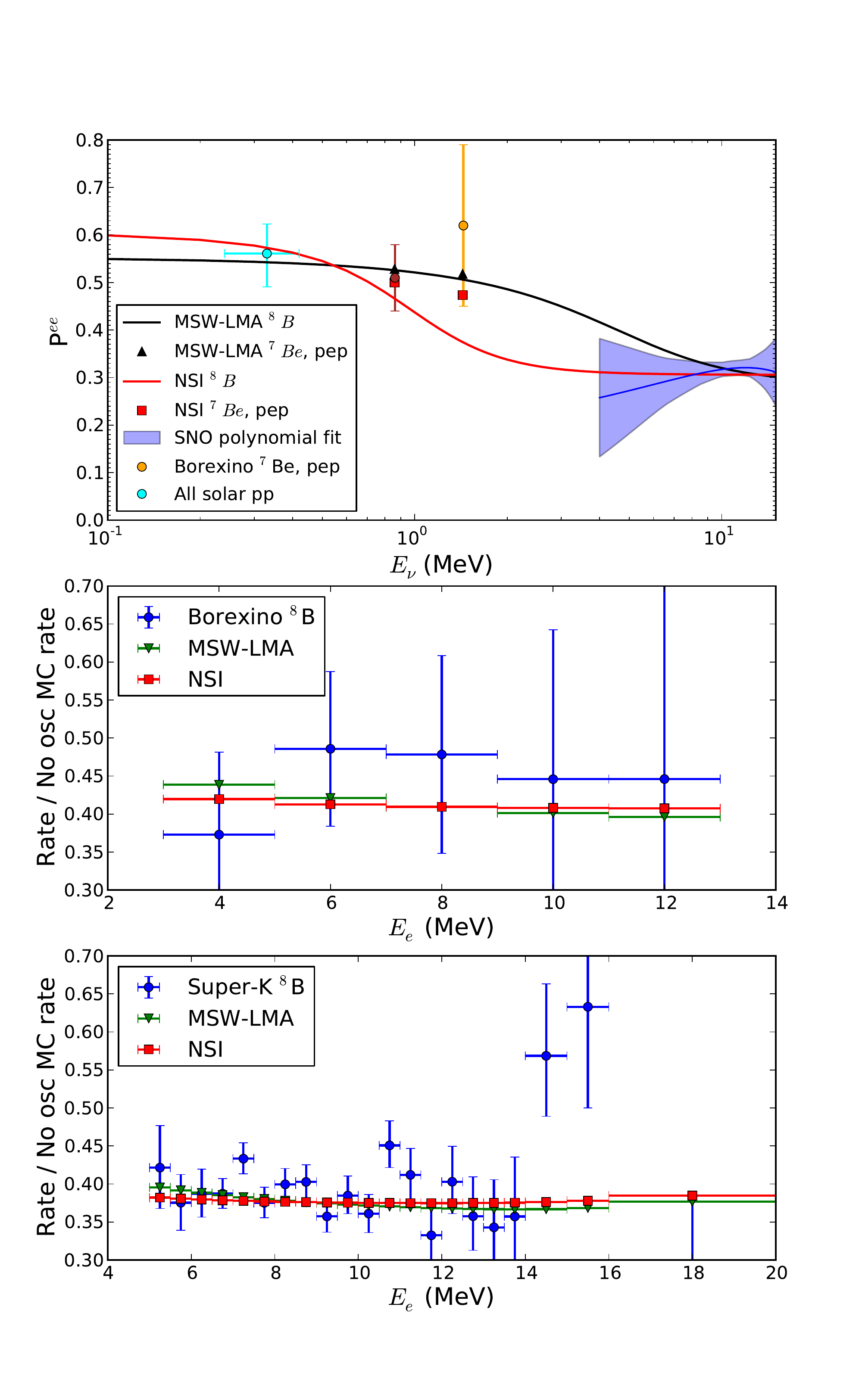}
  \caption{ \label{FIG:lrscalarbestfit}
    (Color online) Best fit for scalar long-range force at $m_{1,0}=0$,
    $\lambda = 1.56 R_{\odot}$, $k_S = 6.73\e{-45}$, $\Delta
    m^2_{21}=7.484\e{-5}$eV$^2, \sin^2\theta_{12}=0.320,
    \sin^2\theta_{13}=0.0239$.  The top plot shows the survival probability as
    a function of incident neutrino energy. The middle shows the best fit's
    predicted event rate in Borexino for each of Borexino's measured electron
    energy bins scaled by the GS98SF2 flux no-oscillation prediction compared
    to Borexino's data, and the bottom shows the same for S-K III's energy bins
    and data.
  }
\end{figure}

\begin{figure}
  \includegraphics[width=0.5\textwidth]{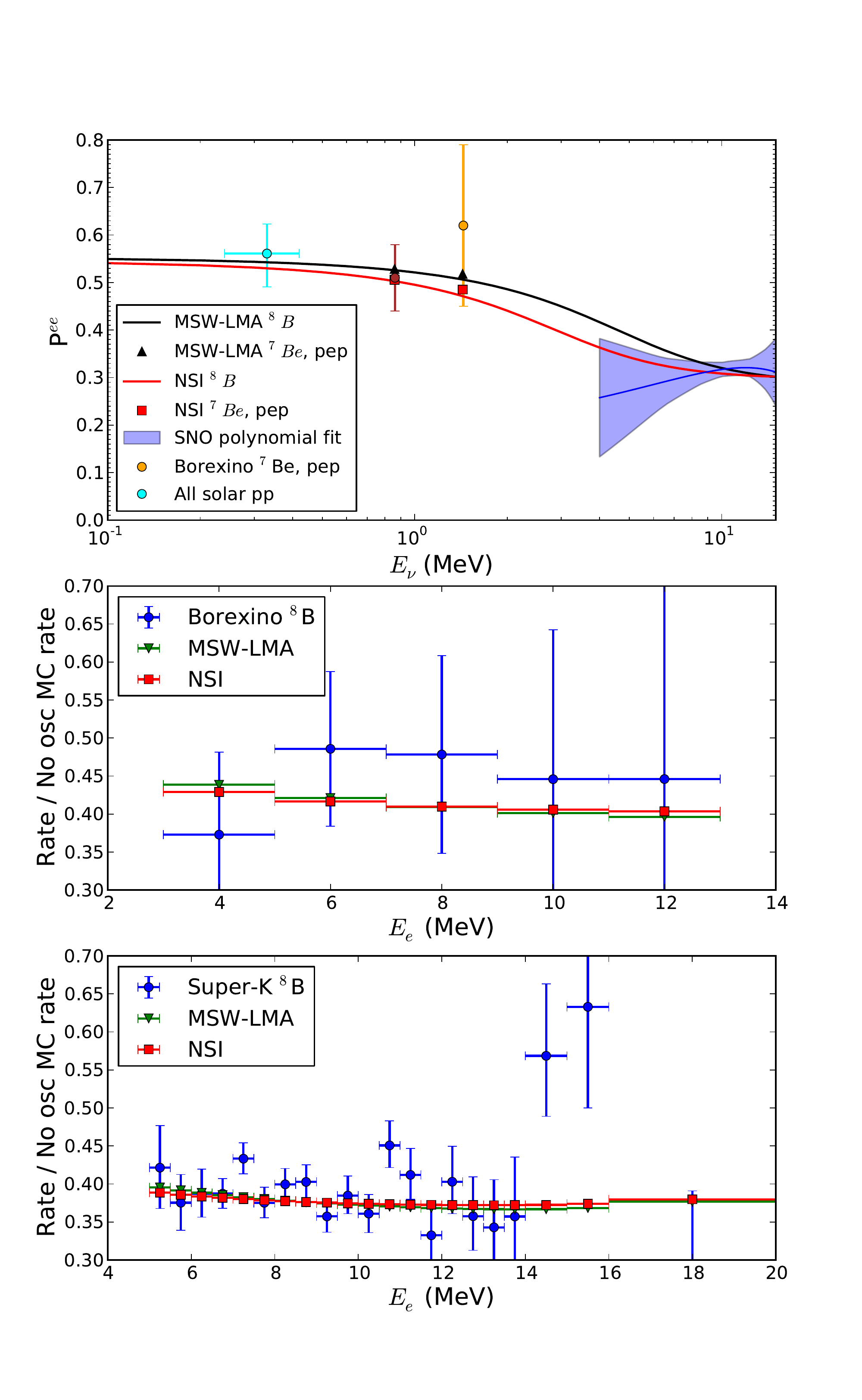}
  \caption{ \label{FIG:lrvectorbestfit}
    (Color online) Best fit for vector long-range force at $\lambda = 16.97
    R_{\odot}$, $k_V = 3.26\e{-54}$, $\Delta m^2_{21}=7.487\e{-5}$eV$^2,
    \sin^2\theta_{12}=0.311, \sin^2\theta_{13}=0.0238$.  The top plot shows the
    survival probability as a function of incident neutrino energy. The middle
    shows the best fit's predicted event rate in Borexino for each of
    Borexino's measured electron energy bins scaled by the GS98SF2 flux
    no-oscillation prediction compared to Borexino's data, and the bottom shows
    the same for S-K III's energy bins and data.
  }
\end{figure}
\end{document}